\documentclass[journal,onecolumn,12pt,draftclsnofoot]{IEEEtran}
\usepackage{amsmath}
\usepackage{amsthm}
\usepackage{amssymb}
\usepackage{ifthen}
\usepackage{graphicx}
\usepackage{subfigure}
\usepackage{color}
\usepackage{balance}
\usepackage{cite}
\usepackage{url}
\newtheorem{definition}{Definition}

\newtheorem{lemma}{Lemma}
\newtheorem{corollary}{Corollary}
\newtheorem{theorem}{Theorem}
\begin{document}
\title{Statistical Age-of-Information Bounds for Parallel Systems: When Do Independent Channels Make a Difference?}
\author{Markus Fidler, Jaya Champati, Joerg Widmer, Mahsa Noroozi\thanks{M. Fidler and M. Noroozi are with the Department of Electrical Engineering and Computer Science, Leibniz University Hannover. J. Champati and J. Widmer are with the IMDEA Networks Institute, Madrid.}}
\maketitle
\begin{abstract}
This paper contributes tail bounds of the age-of-information of a general class of parallel systems and explores their potential. Parallel systems arise in relevant cases, such as in multi-band mobile networks, multi-technology wireless access, or multi-path protocols, just to name a few. Typically, control over each communication channel is limited and random service outages and congestion cause buffering that impairs the age-of-information. The parallel use of independent channels promises a remedy, since outages on one channel may be compensated for by another. Surprisingly, for the well-known case of M$\mid$M$\mid$1 queues we find the opposite: pooling capacity in one channel performs better than a parallel system with the same total capacity. A generalization is not possible since there are no solutions for other types of parallel queues at hand. In this work, we prove a dual representation of age-of-information in min-plus algebra that connects to queueing models known from the theory of effective bandwidth/capacity and the stochastic network calculus. Exploiting these methods, we derive tail bounds of the age-of-information of parallel G$\mid$G$\mid$1 queues. In addition to parallel classical queues, we investigate Markov channels where, depending on the memory of the channel, we show the true advantage of parallel systems. We continue to investigate this new finding and provide insight into when capacity should be pooled in one channel or when independent parallel channels perform better. We complement our analysis with simulation results and evaluate different update policies, scheduling policies, and the use of heterogeneous channels that is most relevant for latest multi-band networks.
\end{abstract}
%
%
\section{Introduction}
The freshness of sensor information that is transmitted via a network to a remote monitor is of vital importance for a variety of applications such as in vehicular communications~\cite{kaul:ageofinformationvehicular, kaul:ageofinformationqueue}, networked feedback control systems~\cite{champati:ageofinformationfeedbackcontrol, ayan:valueofinformation, klugel:aoipenalty}, and cyber-physical systems in general. The age-of-information, or in short age, is a measure that quantifies the information freshness. It is defined as the difference of the current time and the generation time of the latest sensor sample that is available at the monitor. As a consequence, the age has a characteristic saw-tooth shape. It grows linearly with time and it is reset to the network delay whenever a new sample becomes available at the monitor. Today, formulas for the age are known for a catalogue of systems, see the recent surveys~\cite{yates:ageofinformationsurvey, kosta:ageofinformation}. Beyond average and peak age, some works derive the distribution of the age~\cite{inoue:aoisingleserverqueues,champati:ageofinformationgigiqueue,rizk:palmaoi} or bounds of the tail distribution~\cite{champati:ageofinformationmaxplus,noroozi:minplusaoi,noroozi:maxplusadoi}, i.e., of the complementary cumulative distribution function (CCDF).

\begin{figure}
\centering
\includegraphics[width=0.45\linewidth]{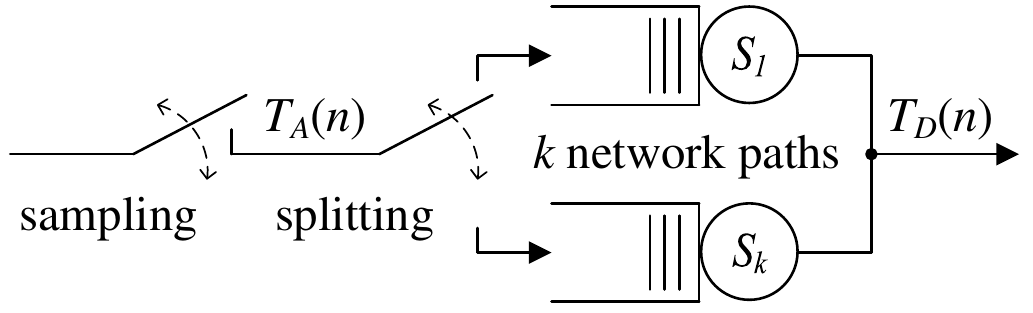}
\caption{Parallel system. A sensor is sampled at times $T_A(n)$ where $n \ge 1$ is the sample index. The samples are transmitted as packets via a network, where the arrival stream is split and transmitted in parallel using $k$ different network paths, denoted as queueing subsystems $S_1, S_2, \dots , S_k$. The resulting departure time stamps $T_D(n)$ are not necessarily in order, since packets may overtake each other on different paths.}
\label{fig:system}
\end{figure}

Current wireless devices increasingly support multiple frequency bands or even multiple technologies, for example WiFi devices operating in 2.4 GHz, 5 GHz, and even 60 GHz channels \cite{perahia2011gigabit}, 5G-NR mobile networks that use sub-6 GHz frequencies as well as the 24-28 GHz millimeter-wave bands \cite{andrews2014will}, or the joint use of a mobile network and WiFi \cite{bajracharya2018lwa}. While current devices typically select only one of those communication options, there is increasing interest in wireless systems that use multiple independent channels simultaneously, to improve performance and robustness. However, for age-of-information in parallel systems, as shown in the example in Fig.~\ref{fig:system}, only a few initial results exist~\cite{raiss:sensinghybridnetworks, altman:foreveryoung,kam:agetransmissionpathdiversity, pan:hybridchannels, bhati:parallelmm1age} and a range of important questions are still open. Here, we focus on age in deployed networks where the first-come first-served (FCFS) discipline is prevalent, there is no control of buffers and queues at intermediate routers, and no feedback is available. Queueing is thus unavoidable and it is usually the most important factor contributing to the age.

The basic proposition of~\cite{bhati:parallelmm1age} is that the age of a parallel system can be expressed as the minimum of the age of the subsystems. This is a substantial difference compared to the commonly known delay metric. It follows that the CCDF of the age is the product of the CCDFs of the age of the subsystems. This holds if the subsystems are statistically independent. The authors of \cite{bhati:parallelmm1age} use this property to derive a solution of the age for a parallel system that consists of independent channels modelled as M$\mid$M$\mid$1 queues. Their main finding is that the average age of a parallel system of two homogeneous M$\mid$M$\mid$1 queues is by a factor of $5/8$ smaller than for a single M$\mid$M$\mid$1 queue. Increasing the parallelism further yields a positive but diminishing return. However, we show that when comparing a system with two homogeneous M$\mid$M$\mid$1 queues to one with a single M$\mid$M$\mid$1 queue but twice the service rate (i.e., the systems have the same overall capacity), there is no gain for parallel channels. In fact, our study of various classical queues\footnote{Kendall's notation A$\mid$S$\mid$1 for a lossless FCFS queue, arrival process A, service process S, and 1 server. A and S can take among others the values M for memoryless, i.e., exponential inter-arrival or service times that form a Poisson process, D for degenerate, i.e., deterministic, E$_l$ for Erlang-$l$, where the special case $l=1$ is exponential, i.e., E$_1 =$ M, and G for general.} such as M$\mid$D$\mid$1, M$\mid$E$_l$$\mid$1, D$\mid$E$_l$$\mid$1 does not show any advantage for parallel systems, which might suggest dismissing parallel systems altogether for the age metric.

\begin{figure}
\centering
\includegraphics[width=0.5\linewidth]{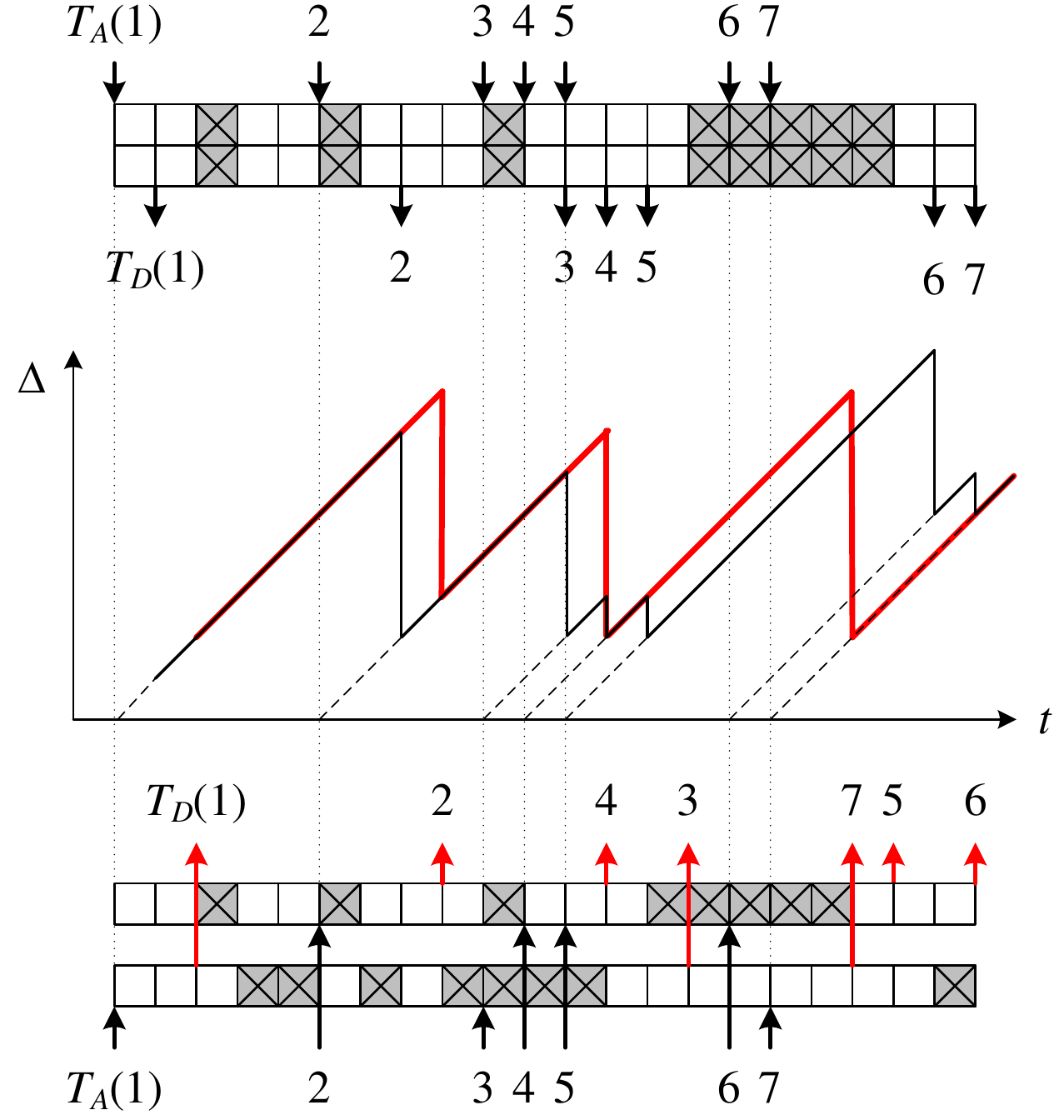}
\caption{Example trajectories of the age of a single on-off channel with capacity 2 (top, black line) compared to two independent parallel on-off channels each with capacity 1 (bottom, red line). Gray, crossed slots mark channel outages.}
\label{fig:aoisawtooth}
\end{figure}

However, real systems may or may not behave like classical queues, and specifically for wireless channels, a more realistic model is the Gilbert-Elliott/two-state Markov channel that provides a certain coherence of the channel over time, and thus it models wireless channel characteristics like outages and memory. In Fig.~\ref{fig:aoisawtooth}, we give an example for the age $\Delta(t)$ of a single on-off channel with capacity 2 (top, black line) and two independent parallel on-off channels each with capacity 1 (bottom, red line). Arrival time-stamps are denoted $T_A(n)$ and departure time-stamps $T_D(n)$. Arrivals and -- in the case of the parallel system -- assignment of packets to channels are random. Packets take two slots and regarding the transmission time the single channel is favorable, see, e.g., packets 1 and 2 that depart from the single channel a unit of time earlier. In the case of parallel channels reordering may happen. The assignment of packet 3 to the second of the parallel channels is unfortunate, but packet 4 provides remedy and makes packet 3 obsolete. On the other hand, packet 7 finds the second channel in on state. It overtakes packets 5 and 6 and improves the age compared to the single channel. We note that in this example, the long outage for the single-channel system results in a higher peak age than the independently occurring outages for the two-channel system.

It is thus vital to investigate how different arrival processes, splitting or scheduling policies, and service processes affect the age-of-information of parallel systems. To this end, we study a broad range of non-trivial random processes, which makes the analysis significantly more difficult. We use models known from effective bandwidths~\cite{kelly:effectivebandwidths, chang:performanceguarantees} and the stochastic network calculus~\cite{chang:performanceguarantees, ciucu:networkservicecurvescaling2, fidler:momentcalculus, jiang:basicstochasticcalculus, fidler:netcalcguide} and use Martingale techniques~\cite{poloczek:servicemartingales, jiang:noteonsnetcalc} to derive tail bounds of arrivals and service to understand the age of relevant systems. The theory includes models of wireless channels, schedulers with random cross-traffic, and tandem networks thereof, as well as basic queueing systems such as the M$\mid$M$\mid$1 queue. Importantly, our analysis of Markov channels shows a true advantage of parallel systems, both for periodic and random updates.

Our contributions are as follows. We formalize age in parallel systems and model the age for variable size packets. We provide a formal mapping of age to min-plus network calculus. This allows us to produce the first comprehensive study of a large variety of parallel systems with a time-varying capacity. Our models include classical queues M$\mid$D$\mid$1, M$\mid$E$_l$$\mid$1, D$\mid$E$_l$$\mid$1, and also more complex Markov channels and further G$\mid$G$\mid$1 systems that can often more accurately represent the behavior of realistic wireless channels, such as on-off channels or systems that adapt the modulation and coding schemes (MCS).

Our analysis provides new insights into how to design and dimension wireless networks with age constraints, for example, in IoT networks. We observe that the age of parallel systems behaves quite differently from the delay. The advantage of parallel independent channels with respect to age is largely due to overtaking of packets, which makes the passed packet obsolete, and thus it differs from the delay. As a consequence, round-robin scheduling compared to the random splitting of Poisson arrivals (both with weights assigned according to the respective channel capacity) improves the delay but has little effect on the age. With round-robin scheduling we achieve load-balancing but loose independence, i.e., a burst of arrivals causes two albeit smaller bursts of arrivals on the parallel paths and less overtaking. Importantly, the benefit of parallel systems is not self-evident and depends on the statistics of the service. We find that the harsher the environment, i.e., with more channel outages and memory, the more beneficial are independent parallel systems. Adding extra capacity to a single channel helps less and less with increasing channel memory, since long off periods of the channel dominate the age. In fact, a mix of very heterogeneous parallel channels in terms of rate and reliability (e.g., 5G-NR FR1 at sub-6 GHz and millimeter-wave FR2) can be useful. If a channel is close(r) to a deterministic channel, the age is generally lower and the precise choice of the optimal update rate at the transmitter brings advantages. Therefore, our results in Sec.~\ref{sec:heterogeneouschannels} suggest that for scheduling on parallel channels, one should first consider the optimal update rate of the 'good' channel and then add remaining update messages on the more bursty channel (where the actual update rate on the bursty channel is less critical).

The remainder of this paper is organized as follows: In Sec.~\ref{sec:parallelsystems} we formalize the age of parallel systems and show initial results for parallel M$\mid$M$\mid$1 queues. Sec.~\ref{sec:minplusmodel} transforms the packet time-stamp model used in the age-of-information literature into cumulative bit processes that are functions of time. This makes a formal connection with the min-plus algebra of the network calculus and opens up a variety of models of systems with a time-varying capacity. As a by-product, the notion of age is extended to variable size packets. In Sec.~\ref{sec:statisticalagebounds} we derive statistical bounds of the age of G$\mid$G$\mid$1 systems. Compared to~\cite{noroozi:minplusaoi}, we take advantage of statistical independence of arrivals and service and use Doob's Martingale inequality to tighten the bounds. Our numerical evaluation of the age of parallel systems is presented in Sec.~\ref{sec:classicalqueues} for classical queues and in Sec.~\ref{sec:markovchannels} for Markov channels. In Sec.~\ref{sec:relatedworks} we discuss further related works and Sec.~\ref{sec:conclusions} provides brief conclusions. Proofs can be found in Sec.~\ref{sec:appendix}.
%
%
\section{Age of Parallel Systems}
\label{sec:parallelsystems}
We investigate the system in Fig.~\ref{fig:system} comprising a sensor that is connected via a network to a remote monitor. At time $T_A(n) > 0$ for $n \in \mathbb{N}$ the sensor is sampled for the $n$th time, the sample is encapsulated in a packet, and transmitted to the network. Hence, $T_A(n)$ is the time-stamp of arrival to the network. Similarly, we denote the departure time-stamp from the network to the monitor $T_D(n)$. For causality, $T_D(n) \ge T_A(n)$ for all $n$. Time is continuous $t \in \mathbb{R}_{\ge 0}$ and the processes $T_A(n)$ and $T_D(n)$ are non-decreasing, i.e., packets are labelled in ascending order of their time-stamps. By convention there are no arrivals at $t=0$ and for convenience we define $T_A(0) = 0$ and $T_D(0) = 0$. For now all packets have unit size. Later in Sec.~\ref{sec:minplusmodel} we consider variable-size packets.

The network offers $k$ disjoint network paths denoted subsystems $S_i$ for $i \in \{1,2,\dots,k\}$ that can be used in parallel. The arrivals are divided among the $k$ subsystems according to a defined scheduling policy forming the arrival processes $T_{A_i}(n_i)$ that have the respective departure processes $T_{D_i}(n_i)$, where $n_i \in \mathbb{N}$ is the consecutive packet count at each of the subsystems $i$. We assume lossless FCFS transmission, i.e., each individual subsystem delivers packets in order. Packets at different subsystems can incur, however, different delays and may overtake each other. The processes $T_{A_i}(n_i)$ and $T_{D_i}(n_i)$ can be superposed in ascending order of the arrival time-stamps to form the processes $T_A(n)$ and $T_D(n)$~\cite[Eq. 3.2]{liebeherr:duality}. The mathematical formalism is, however, complex since in order aggregation involves sorting. For the following investigation of the age-of-information, this step is not required.

A definition of the age-of-information, or in short age, at time $t \ge T_D(1)$ is given, e.g., in~\cite{champati:ageofinformationmaxplus}
\begin{equation}
\Delta(t) = t - \max_{n \ge 1} \{ T_A(n): T_D(n) \le t \} .
\label{eq:age}
\end{equation}
Likewise, $\Delta_i(t) = t - \max_{n_i \ge 1} \{ T_{A_i}(n_i): T_{D_i}(n_i) \le t \}$ holds for each of the subsystems on its own. We note that~\eqref{eq:age} considers each tuple $(T_A(n), T_D(n))$ individually, so that any reordering of the tuples gives the same age. Hence, ordering the tuples in ascending order is not required and we can write that
\begin{equation*}
\Delta(t) = t - \max_i \Bigl\{ \max_{n_i} \{ T_{A_i}(n_i): T_{D_i}(n_i) \le t \} \Bigr\} ,
\end{equation*}
which can be rearranged as
\begin{equation*}
\Delta(t) = \min_i \Bigl\{ t - \max_{n_i} \{ T_{A_i}(n_i): T_{D_i}(n_i) \le t \} \Bigr\} .
\end{equation*}
By insertion of $\Delta_i(t) = t - \max_{n_i} \{ T_{A_i}(n_i): T_{D_i}(n_i) \le t \}$ this proves the proposition made in~\cite{bhati:parallelmm1age} that
\begin{equation}
\Delta(t) = \min_i \{ \Delta_i(t) \} ,
\label{eq:ageparallel}
\end{equation}
i.e., the age of a parallel system is conveniently obtained as the minimal age of the subsystems.

We will focus on systems where the age varies due to random arrivals and random service. Considering the CCDF of $\Delta(t)$ we have with~\eqref{eq:ageparallel} that
\begin{equation*}
\mathsf{P}[\Delta(t) > x] = \mathsf{P} \Biggl[\bigcap_i \, \Delta_i(t) > x \Biggr] ,
\end{equation*}
and for statistically independent subsystems
\begin{equation}
\mathsf{P}[\Delta(t) > x] =  \prod_i \mathsf{P} [ \Delta_i(t) > x] .
\label{eq:ageindependent}
\end{equation}

This makes it possible to study different transmission policies like load-balancing or replication depending on the goal of reducing the age or increasing the reliability.

For an illustration of this basic result, we consider the age-of-information of the well-known M$\mid$M$\mid$1 queue. For this case, the CCDF of the age is known in closed form~\cite{inoue:aoisingleserverqueues}. Since random splitting of a Poisson process results in independent Poisson arrival processes the precondition of~\eqref{eq:ageindependent} is satisfied and the age of parallel M$\mid$M$\mid$1 queues is readily obtained thereof~\cite{bhati:parallelmm1age}.

In a numerical example, we compare two systems: one with a single and one with two parallel M$\mid$M$\mid$1 queues. In the case of the parallel system, the queues are independent and packets may overtake each other. This is expected to benefit the age: if packets are stuck in congestion in one of the queues, packets may nevertheless be served timely from the other queue and avoid excessive growth of the age.

\begin{figure}
\centering
\includegraphics[width=0.51\linewidth]{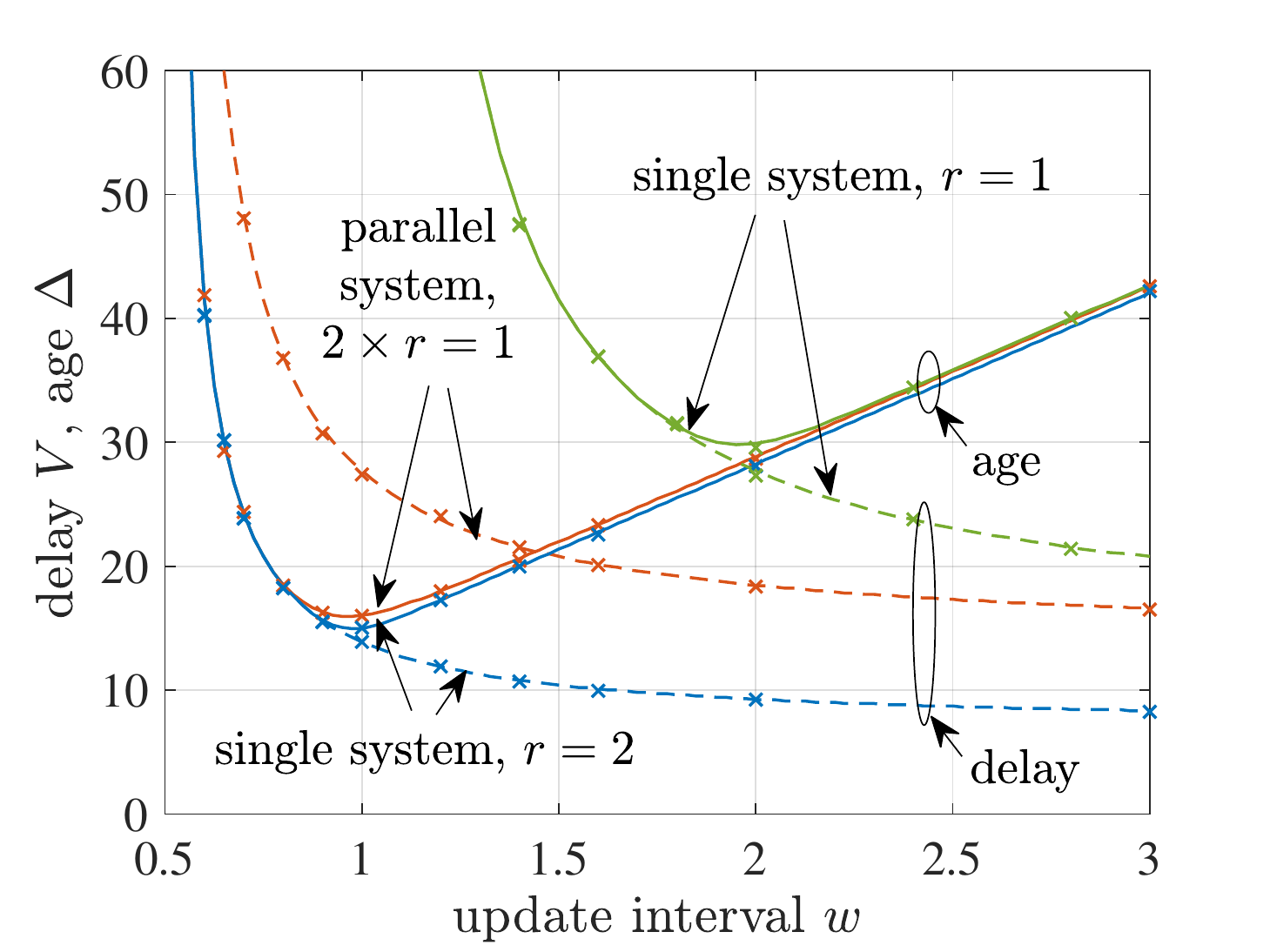}
\caption{Single versus parallel M$\mid$M$\mid$1 queues. We show the CCDF at $\varepsilon=10^{-6}$. Lines are analytical results and the markers simulation results. While the parallel system improves the age compared to the single system with mean service rate $r=1$, it is outperformed by the single system with rate $r=2$.}
\label{fig:mm1_exact}
\end{figure}

In Fig.~\ref{fig:mm1_exact} we show tail bounds of the age that are exceeded at most with probability $\varepsilon = 10^{-6}$. We also include tail bounds of the delay that are obtained from the delay distribution of the M$\mid$M$\mid$1 queue see, e.g.,~\cite{bolch:queueingnetworks}. The analytical results are represented by lines and the markers in Fig.~\ref{fig:mm1_exact} are $(1-10^{-6})$-quantiles obtained from the simulation of $10^9$ packets for each data point. Parameter $w$ on the x-axis is the update interval that is the reciprocal of the mean rate of the external Poisson arrival process before splitting.

First, we consider a single M$\mid$M$\mid$1 queue with mean service rate $r=1$. We notice that the CCDF of the age is minimized at a mean arrival rate of $1/w \approx 0.5$. For smaller $w$ the age increases due to congestive delays, whereas for larger $w$ the delay decreases but the age increases due to longer time gaps between updates.

For the parallel system consisting of two M$\mid$M$\mid$1 queues each with mean rate $r=1$ the minimal age improves by almost a factor of two. It is attained for an arrival rate of $1/w \approx 1$, that is the stability limit of the single queue but feasible in the parallel system due to load-balancing.

Lastly, we include results for a single M$\mid$M$\mid$1 queue with mean service rate $r=2$, i.e., the service rate equals the sum of the service rates available in the parallel system. If compared to this system, the parallel system turns out to be inferior. Its delay is significantly larger, and its age slightly. A certain increase is explained by the service time if a packet is served at mean rate $r=1$ by one of the parallel queues compared to $r=2$ by the single system. The simulation results as well as an analysis of the average age confirm the same observation.

This raises the question whether the advantage of parallel systems observed in~\cite{bhati:parallelmm1age} is only a consequence of the increase in capacity when adding further subsystems. Without the availability of other models of parallel systems, it remains unanswered whether single systems of equal total capacity will generally perform better than parallel systems, or if there are cases where the independence of the subsystems of a parallel system can help to outperform the capacity-equivalent single system. The theory that we develop in Sec.~\ref{sec:minplusmodel} and~\ref{sec:statisticalagebounds} includes a variety of models that allow to reason about these questions.
%
%
\section{Min-plus Model of Age}
\label{sec:minplusmodel}
In this section we transform the definition of age~\eqref{eq:age} from packet time-stamps $T_A(n)$ to cumulative arrival functions $A(t)$ denoting the cumulative number of bits that arrived until time $t$. The reason for this change of notation is that it allows for a variety of system models including schedulers, wireless channels, and networks thereof. For example, a wireless channel's variations over time are not related to specific packet indices and are most naturally modelled by a service process $S(\tau,t)$ expressing the service available between times $\tau$ and $t$.

The notion of service processes, that will be defined in~\eqref{eq:serviceprocess}, leads to a min-plus system theory for which the stochastic network calculus provides a toolbox of methods for performance analysis, e.g.,~\cite{chang:performanceguarantees, ciucu:networkservicecurvescaling2, fidler:momentcalculus, jiang:basicstochasticcalculus, fidler:netcalcguide}. Age-of-information bounds are derived for the first time in~\cite{noroozi:minplusaoi} in the min-plus network calculus, however, without formal reference to the original definition of age~\eqref{eq:age}. We close this gap and provide the proof below.

First, in Sec.~\ref{sec:continuousspace} we extend the packet time-stamp model to packets of variable size. We make the connection with the min-plus model in Sec.~\ref{sec:mappingminplus}, where we use that cumulative functions $A(t)$ are inverse functions of packet time-stamp processes $T_A(n)$, i.e., $A(t)$ is the reflection of $T_A(n)$ at the $45^{\circ}$ line and vice versa. The reader may skip to the main result~\eqref{eq:ageminplus} that is the age expressed by cumulative arrival and departure functions $A(t)$ and $D(t)$, respectively. Then, Sec.~\ref{sec:minplusservice} presents how $D(t)$ is related to $A(t)$ given the system's service process $S(\tau,t)$.
%
%
\subsection{Continuous-space model}
\label{sec:continuousspace}
We use the extension of the discrete-space model of arrivals $T_A(n)$ for $n \in \mathbb{N}$ to continuous-space $n \in \mathbb{R}_{> 0}$ worked out by~\cite{liebeherr:duality}. The continuous-space model is a generalization that extends previous results to packets of variable size. The change of model is not strictly required for this work (we can derive the following results similarly if we depart from discrete functions $T_A(n)$), but it achieves consistency.

The discrete model of packet time-stamps $T_A(n)$ can be linked to variable size packets using an additional function $l(n)$ that defines the corresponding sequence of packet lengths~\cite{chang:performanceguarantees}. These two functions, time-stamps and packet lengths, can be conveniently integrated into a bit-level representation $T_A(n)$, where $n$ denotes bits instead of packets, or into a fluid-flow model, where bits are infinitely divisible and $n \in \mathbb{R}_{> 0}$.

For an example consider the discrete arrival time-stamps $T_A(n) = \{1.5,3,4\}$ of packets $n \in \{1,2,3\}$ with lengths $l(n) = \{500, 700, 200 \}$. In the continuous-space model this translates to $T_A(0) = 0$ by our convention, $T_A(n) = 1.5$ for $n \in (0,500]$ for packet one, $T_A(n) = 3$ for $n \in (500,1200]$ for packet two, and $T_A(n) = 4$ for $n \in (1200,1400]$ for packet three. As can be seen, $T_A(n)$ is left-continuous. We make this convention different from~\cite{liebeherr:duality} as it will fit more naturally with the following derivation of the age. The departure function $T_D(n)$ is defined in the same way.

When using continuous instead of discrete functions we replace the operators $\max$ by $\sup$ and $\min$ by $\inf$, respectively. It is straightforward to verify that the definition of age~\eqref{eq:age} as well as the age of parallel systems~\eqref{eq:ageparallel} remain otherwise unchanged, when we switch to the continuous-space model.
%
%
\subsection{Mapping to min-plus algebra}
\label{sec:mappingminplus}
Let $T_A(n)$ and $T_D(n)$ be continuous (or discrete) arrival and departure time functions. We define the cumulative function $A(t) \ge 0$ that counts the cumulative number of bits (or packets) that arrived until time $t \ge 0$. Clearly, $A(t)$ is non-decreasing and by our convention $A(0) = 0$. Shorthand notation $A(\tau,t) = A(t) - A(\tau)$ denotes the arrivals in an interval $(\tau,t]$ for $t \ge \tau \ge 0$. We use a fluid-flow model and make the convention that the cumulative functions are right-continuous, i.e., a bit (or packet) that arrives at $t$ is included in $A(t)$. Similarly, $D(t)$ denotes the departures. It holds that
\begin{equation}
A(t) = \sup \{ \nu : T_A(\nu) \le t \} =: T_A^{\uparrow}(t),
\label{eq:upperpseudoinverse}
\end{equation}
and
\begin{equation}
T_A(n) = \inf \{ \tau : A(\tau) \ge n \} =: A^{\downarrow}(n),
\label{eq:lowerpseudoinverse}
\end{equation}
and accordingly for $D(t)$ and $T_D(n)$. The functions $T_A^{\uparrow}(t)$ and $A^{\downarrow}(n)$ are the upper and lower pseudo-inverse of $T_A(n)$ and $A(t)$, respectively. Pseudo-inverse instead of inverse functions have to be used since the functions have steps and plateaus. Generally the upper pseudo-inverse is not smaller than the lower pseudo-inverse, the pseudo-inverse of a non-decreasing function is non-decreasing, the lower pseudo-inverse is left-continuous, and the upper-pseudo inverse is right-continuous. Pseudo-inverses establish an isomorphism of packet time-stamp functions and cumulative arrival functions. For a complete elaboration see~\cite{liebeherr:duality}. We use this duality to formally map the definition of age~\eqref{eq:age} to a corresponding expression in the min-plus network calculus.

For $T_A(n)$ non-decreasing, the expression of the age~\eqref{eq:age} can be expanded as
\begin{equation}
\Delta(t) = t - T_A(n^*) ,
\label{eq:ageexpanded}
\end{equation}
where
\begin{equation}
n^*(t) = \sup \{ \nu : T_D(\nu) \le t \} .
\label{eq:ageexpandedindex}
\end{equation}
Now,~\eqref{eq:ageexpandedindex} matches the form of~\eqref{eq:upperpseudoinverse}, i.e., $n^*(t) = T_D^{\uparrow}(t) = D(t)$. By insertion into~\eqref{eq:ageexpanded} we have $\Delta(t) = t - T_A(D(t))$ and by substitution of~\eqref{eq:lowerpseudoinverse} for $T_A(D(t)) = A^{\downarrow}(D(t))$ it follows that
\begin{equation*}
\Delta(t) = t - \inf \{ \tau : A(\tau) \ge D(t) \} .
\end{equation*}
We note that $\tau \in [0,t]$ since $A(t) \ge D(t)$. After some reformulation
\begin{equation*}
\Delta(t) = \sup_{\tau \in [0,t]} \{ t - \tau : D(t) - A(\tau) \le 0 \} ,
\end{equation*}
and by substitution of $\delta = t-\tau$
\begin{equation}
\Delta(t) = \sup_{\delta \in [0,t]} \{ \delta : D(t) - A(t-\delta) \le 0\} ,
\label{eq:ageminplus}
\end{equation}
we arrive at an equivalent formulation of the age-of-information in the min-plus network calculus. This proves a proposition that has been used previously in~\cite{noroozi:minplusaoi}. For intuition, at time $t$ the most recent information at the monitor is $D(t)$. Now~\eqref{eq:ageminplus} searches for the largest $\delta$ corresponding to the earliest time $t-\delta$ at which this information has been available and thus has been sent to the network as arrival $A(t-\delta)$.
%
%
\subsection{Min-plus service characterization}
\label{sec:minplusservice}
To evaluate $\Delta(t)$ given in~\eqref{eq:ageminplus} we require a system model that enables estimating the departure function $D(t)$. We characterize the service provided by a transmission system, e.g., a link, a wireless channel, or a scheduler, by a service process $S(\tau,t) \ge 0$, that specifies the amount of service available in an interval $(\tau,t]$. A system has lower service process $S(\tau,t)$ if it holds for all $t \ge 0$ that~\cite[Def. 5.5.1]{chang:dynamicserviceguarantees}
\begin{equation}
D(t) \ge \inf_{\tau \in [0,t]} \{ A(\tau) + S(\tau,t) \} =: A \otimes S(t) .
\label{eq:serviceprocess}
\end{equation}
The operator $\otimes$ is known as min-plus convolution. Eq.~\eqref{eq:serviceprocess} is a model of a G$\mid$G$\mid$1 queue. A simple example of~\eqref{eq:serviceprocess} is a buffered constant rate link with rate $r$ that has service process $S(\tau,t) = r (t-\tau)$.

The concept of service process uses a fluid-flow model. The additional effects that are due to packet boundaries are modelled by a packetizer~\cite{leboudec:networkcalculus, chang:performanceguarantees}. Given packet lengths $l(n)$ with packet index $n \in \mathbb{N}$, let $L(n) = \sum_{\nu=1}^n l(\nu)$ denote the cumulative packet length and $L(0) = 0$. The packetizer takes fluid input $x$ and converts it into packetized output
\begin{equation}
P_L(x) = \max_{n \in \mathbb{N}_0} \bigl\{L(n) 1_{\{L(n) \le x\}} \bigr\},
\label{eq:packetizer}
\end{equation}
i.e., it delays the bits of a packet until the packet is completed. Above, $1_{\{.\}}$ is the indicator function that is one if the argument is true and zero, otherwise. It follows that $x \ge P_L(x) \ge x -l_{\max}$, where $l_{\max} = \max_{n \in \mathbb{N}} \{l(n)\}$ is the maximal packet length. A function $A(t)$ is packetized if $A(t)=P_L(A(t))$. Given packetized arrivals, the effects that are due to packetization of the departures can be conveniently integrated into the lower service process, i.e., the tandem of the fluid system and the packetizer offers service process~\cite[Eq. 4]{noroozi:minplusaoi}
\begin{equation}
S_{P_L}(\tau,t) = [S(\tau,t) - l_{\max}]_+ ,
\label{eq:servicepacketizer}
\end{equation}
where $[x]_+ = \max \{0,x\}$ is the non-negative part.
%
%
\section{Statistical Age Bounds}
\label{sec:statisticalagebounds}
Th.~\ref{th:aoi} in this section gives our main result, which is a statistical tail bound on the age-of-information of systems with general arrival and service processes, where the service times can be correlated in time. While Th.~\ref{th:aoi} is derived for single systems, the form of the result $\mathsf{P}[\Delta(t) > x]$ is compatible with~\eqref{eq:ageindependent} and thus it extends immediately to parallel systems.

Starting point of our derivation is the min-plus formulation of age~\eqref{eq:ageminplus} where we substitute the concept of service process~\eqref{eq:serviceprocess} and packetizer~\eqref{eq:servicepacketizer}. We view arrivals $A(t)$ and service $S(\tau,t)$ as random processes and characterize these processes by statistical envelope functions. The envelopes are determined by two parameters $(\rho, \varepsilon)$ defined in Def.~\ref{def:envelopes}. The parameters of the envelope functions are then inserted into Th.~\ref{th:aoi}. We will specify the envelope parameters of a variety of classical queues and Markov channels in Sec.~\ref{sec:classicalqueues} and~\ref{sec:markovchannels}, respectively. Most importantly, the reader who skips the mathematical details of this section should remember that for the rest of this work we only need the envelope parameters $(\rho, \varepsilon)$ to apply Th.~\ref{th:aoi}.

A tail bound for single systems that is related to Th.~\ref{th:aoi} has also been derived in~\cite{noroozi:minplusaoi}. In this paper, we substantially tighten the bound compared to~\cite{noroozi:minplusaoi} as it turned out that tightness was more of a concern when using~\eqref{eq:ageindependent} for parallel systems. The improvement of Th.~\ref{th:aoi} is achieved by taking the statistical independence of arrivals and service into account. Secondly, we use envelopes that are derived via Doob's Martingale inequality and that are tighter than those obtained from union and Chernoff bound in~\cite{noroozi:minplusaoi}.

In detail, we use a basic rate envelope with an overflow or underflow profile that is typically an exponential function. More general envelope functions are feasible, though not needed for this work. Envelopes originated as a model of arrivals and service under the names exponentially bounded burstiness and exponentially bounded fluctuation in~\cite{yaron:exponentiallyboundedburstiness} and~\cite{lee:exponentiallyboundedfluctuation}, respectively. Successive works~\cite{cruz:qosmanagement, li:effectivebandwidthcalculus2, yin:generalizedstochasticallyboundedburstiness, ciucu:networkservicecurvescaling2} derived envelopes for entire sample paths that are of the type of Def.~\ref{def:envelopes}. We have chosen a model with separate envelopes for the arrivals and service as this achieves modularity. Envelope functions that satisfy Def.~\ref{def:envelopes} are known for many relevant system models including wireless channels, schedulers, and multi-node networks thereof~\cite{fidler:netcalcguide}. Our new theorem Th.\ref{th:aoi} for the age-of-information is built to encompass all of these models, making it a versatile tool for analyzing many different systems. Proofs can be found in Sec.~\ref{sec:appendix}.
\begin{definition}[Statistical Envelopes]
\label{def:envelopes}
The service has lower envelope rate $\rho_S > 0$ with underflow profile $\varepsilon_S(b) \ge 0$ and burstiness parameter $b \ge 0$ if for all $t \ge 0$
\begin{equation*}
\mathsf{P}[\exists \tau \in [0,t]: S(\tau,t) < [\rho_S (t-\tau) - b]_+] \le \varepsilon_S(b) .
\end{equation*}
The arrivals have upper envelope rate $\rho_A > 0$ with overflow profile $\overline{\varepsilon}_A(b) \ge 0$ and burstiness parameter $b \ge 0$ if for all $t \ge 0$
\begin{equation*}
\mathsf{P}[\exists \tau \in [0,t]: A(\tau,t) > \rho_A (t-\tau) + b] \le \overline{\varepsilon}_A(b) ,
\end{equation*}
and lower envelope with $l_{\min} > 0$ denoting the minimal packet size and underflow profile $\underline{\varepsilon}_A(u)$ with parameter $u \ge 0$ if for all $t \ge 0$
\begin{equation*}
\mathsf{P}[\exists \tau \ge t: A(t,\tau) < 1_{\{\tau-t>u\}} l_{\min}] \le \underline{\varepsilon}_A(u).
\end{equation*}
All $\varepsilon$ are non-negative, non-increasing, continuous functions.
\end{definition}
\begin{corollary}[Service Process and Packetizer]
\label{cor:packetizer}
For a system with packetizer~\eqref{eq:servicepacketizer} and maximum packet length $l_{\max} > 0$ it follows from Def.~\ref{def:envelopes} for $\rho_S > 0$, $\varepsilon_S(b) \ge 0$, $b \ge 0$, and all $t \ge 0$ that
\begin{equation*}
\mathsf{P}[\exists \tau \in [0,t]: S_{P_L}(\tau,t) < [\rho_S (t-\tau) - l_{\max} - b]_+] \le \varepsilon_S(b) .
\end{equation*}
\end{corollary}
\begin{theorem}[Statistical Age Bound]
\label{th:aoi}
With Def.~\ref{def:envelopes} and Cor.~\ref{cor:packetizer}, $\rho_A \le \rho_S$ for stability, and assuming independence of arrivals and service, it holds for $t \ge x$ and $x \ge l_{\max}/\min\{\rho_S,\rho_S'\}$ that
\begin{align*}
\mathsf{P}[\Delta(t) > x] \le & 1 - [1-\overline{\varepsilon}_A]_+ \ast [1-\varepsilon_S]_+ (\rho_S x - l_{\max}) \\
+ & 1 - [1-\underline{\varepsilon}_A]_+ \ast [1-\varepsilon_T]_+ (x - l_{\max}/\rho_S') ,
\end{align*}
where $\varepsilon_T(u) = \varepsilon_S'(\rho_S' u)$ and $(\rho_S', \varepsilon_S')$ is a lower service envelope, and the operator $\ast$ is the Stieltjes convolution defined as $f \ast g(x) = \int_0^x f(x-y) dg(y)$.
\end{theorem}
Th.~\ref{th:aoi} specifies an upper bound of the age $x$ that is exceeded at most with probability \mbox{$\mathsf{P}[\Delta(t) > x]$.} The probability is composed of two terms corresponding to the two lines in Th.~\ref{th:aoi}:
\begin{enumerate}
    \item The probability that the age exceeds $x$ due to queueing in the network. Here, the term $\rho_S x$ is the amount of backlog that can be served at rate $\rho_S$ in $x$ units of time. The backlog may be caused by bursty arrivals, expressed by $\overline{\varepsilon}_A$, and service outages, expressed by $\varepsilon_S$.
    \item The probability that the age exceeds $x$ due to idle waiting for new updates. The waiting time $x$ is composed of the inter-arrival time of updates, expressed by $\underline{\varepsilon}_A$, and a service time including potential service outages, expressed by $\varepsilon_T$.
\end{enumerate}
Lastly, the effect of the maximal packet size $l_{\max}$, where $l_{\max}/\rho_S'$ is the effective transmission time of the packet, is included.

We note that the upper bound in Th.~\ref{th:aoi} may take values greater than one. Without explicit mention, we generally truncate bounds of probabilities so that they are in $[0,1]$.

The following lemma will be useful to compute the Stieltjes convolution. We note that convolution is commutative.
\begin{lemma}[Stieltjes convolution]
\label{lem:convolution}
Given $\varepsilon_1(x) = \alpha e^{-\upsilon_1 x}$ and $\varepsilon_2(x) = e^{-\upsilon_2 x}$ with $\alpha, \upsilon_1, \upsilon_2 > 0$. For $x \ge [\ln(\alpha)]_+/\upsilon_1$ we have
\begin{align*}
1 - [1-\varepsilon_1]_+ \ast (1-\varepsilon_2) (x)
=
\begin{cases}
(1 + \alpha \upsilon x) e^{-\upsilon x}, & \text{ if } \alpha \le 1, \upsilon_1 = \upsilon_2 =: \upsilon, \\
e^{-\upsilon_2 x} + \frac{\alpha \upsilon_2}{\upsilon_1-\upsilon_2} \bigl(e^{-\upsilon_2 x} - e^{-\upsilon_1 x}\bigr), & \text{ if } \alpha \le 1, \upsilon_1 \neq \upsilon_2, \\
\alpha(1 - \ln \alpha + \upsilon x) e^{-\upsilon x}, & \text{ if } \alpha > 1, \upsilon_1 = \upsilon_2 =: \upsilon, \\
\frac{1}{\upsilon_1-\upsilon_2}\bigl(\alpha^{\upsilon_2/\upsilon_1} \upsilon_1 e^{-\upsilon_2 x} - \alpha \upsilon_2 e^{-\upsilon_1 x}\bigr), & \text{ if } \alpha > 1, \upsilon_1 \neq \upsilon_2. \\
\end{cases}
\end{align*}
\end{lemma}

In the following presentation of results, we will also evaluate the delay for comparison with the age. For FCFS systems, the virtual delay is defined as
\begin{equation}
V(t) = \inf_{\tau \ge 0} \{ \tau : A(t) - D(t+\tau) \le 0\} .
\label{eq:virtualdelay}
\end{equation}
The definition of delay is referred to as virtual since it is not conditioned on a packet arrival. It holds that
\begin{equation*}
\mathsf{P}[V(t) > x] \le 1 - [1-\overline{\varepsilon}_A]_+ \ast [1-\varepsilon_S]_+ (\rho_S x) .
\end{equation*}
The derivation is similar to the first part of the proof of Th.~\ref{th:aoi}, except that the packetizer is omitted as it does not increase the delay at the system where it is appended~\cite[Th. 1.7.1]{leboudec:networkcalculus}.

Next, we state two corollaries that cover the special cases of deterministic service and deterministic arrivals, respectively.
\begin{corollary}[Age Bound for Constant Rate Service]
\label{cor:constantrateservice}
Consider the assumptions of Th.~\ref{th:aoi} and use constant rate service $S_{P_L}(\tau,t) = [r (t-\tau) - l_{\max}]_+$ with rate $r > 0$ and maximum packet size $l_{\max} > 0$. With $\rho_A \le r$ for stability it holds for $t \ge x$ and $x \ge l_{\max}/r$ that
\begin{equation*}
\mathsf{P}[\Delta(t) > x] \le \overline{\varepsilon}_A(r x - l_{\max}) + \underline{\varepsilon}_A(x - l_{\max}/r) .
\end{equation*}
\end{corollary}
\begin{corollary}[Age Bound for Periodic Arrivals]
\label{cor:periodicarrivals}
Consider the assumptions of Th.~\ref{th:aoi} and use periodic arrivals $A(t) = l \lfloor (t+o)/w \rfloor $ with constant packet length $l > 0$, constant inter-arrival time $w > 0$, and time offset $o \in [0,w)$. With $l/w \le \rho_S$ for stability it holds for $t \ge x$, and $x \ge w + l/\min\{\rho_S,\rho_S'\}$ that
\begin{align*}
\mathsf{P}[\Delta(t) > x] \le & \varepsilon_S(\rho_S (x - [l/\rho_S - (t-x+o) \bmod w]_+) - l) \\
+ & \varepsilon_S'(\rho_S' (x - (w - (t-x+o) \bmod w)) - l) .
\end{align*}
\end{corollary}
Cor.~\ref{cor:periodicarrivals} takes the phase of periodic arrivals into account. A consequence is that the bound is not a continuous function in $t$ and $x$. The phase will be important when we consider parallel systems later.
%
%
\section{Parallel Classical Queues}
\label{sec:classicalqueues}
In this and the following section we will use Th.~\ref{th:aoi} to derive tail bounds of the age of a variety of parallel systems. In this section we start with classical queueing systems before we consider more complicated Markov channels in Sec.~\ref{sec:markovchannels}. The concepts that we introduce here will be used in Sec.~\ref{sec:markovchannels} again. The results of this section give insights into the performance of parallel systems where the subsystems' service processes are not correlated in time.

In order to use Th.~\ref{th:aoi} we need to determine the envelope parameters of arrivals and service according to Def.~\ref{def:envelopes}. We use the moment generating function (MGF) $\mathsf{E} [e^{\theta A(t)}]$ where $\theta$ is a free parameter to characterize the distribution of random arrival processes $A(t)$. For the simple case of Lévy arrival processes, that are processes with stationary independent increments, it holds that
\begin{equation}
\frac{1}{\theta t}\mathsf{E} \bigl[e^{\theta A(t)}\bigr] = \frac{1}{\theta}\mathsf{E} \bigl[e^{\theta A(1)}\bigr] =: \rho_A(\theta)
\label{eq:mgfenvelope}
\end{equation}
which is independent of $t>0$ and expressed as envelope rate $\rho_A(\theta)$ for $\theta > 0$. The envelope rate $\rho_A(\theta)$ increases with $\theta$ and lies between the mean and the peak arrival rate (possibly infinity). For a more detailed explanation of the nature of the parameter $\theta$ see~\cite{kelly:effectivebandwidths}.

An upper arrival envelope that satisfies Def.~\ref{def:envelopes} is derived for Lévy processes in~\cite[Lem. 6]{jiang:noteonsnetcalc} using Doob's martingale inequality. The envelope has parameter $\rho_A(\theta)$~\eqref{eq:mgfenvelope} for $\theta > 0$ with overflow profile
\begin{equation}
\overline{\varepsilon}_A(b) = e^{-\theta b} .
\label{eq:overflowprofile}
\end{equation}
Here, $\theta$ determines the speed of the decay of the overflow profile $\overline{\varepsilon}_A$ of the envelope rate $\rho_A$, e.g., if we choose a larger parameter $\theta$, we obtain a larger envelope rate that has, however, a smaller probability of overflow.

A lower service envelope that satisfies Def.~\ref{def:envelopes} can be derived along the same line as~\cite[Lem. 6]{jiang:noteonsnetcalc}. For completeness we include the derivation in the appendix.
The service process is characterized by the Laplace transform that is the negative MGF. For Lévy service processes it holds for all $\tau \ge 0$ and $t > 0$ that
\begin{equation}
-\frac{1}{\theta t}\mathsf{E} \bigl[e^{-\theta S(\tau,\tau+t)}\bigr] = -\frac{1}{\theta}\mathsf{E} \bigl[e^{-\theta S(0,1)}\bigr] =: \rho_S(\theta) .
\label{eq:laplaceenvelope}
\end{equation}
The envelope rate $\rho_S(\theta)$ for $\theta > 0$ decreases with $\theta$ and lies between the mean and the minimal service rate (possibly zero). For any $\theta > 0$, the envelope rate $\rho_S(\theta)$ satisfies Def.~\ref{def:envelopes} with underflow profile
\begin{equation}
\varepsilon_S(b) = e^{-\theta b} .
\label{eq:serviceunderflowprofile}
\end{equation}
We note that $\theta$ in~\eqref{eq:mgfenvelope},~\eqref{eq:overflowprofile} and $\theta$ in~\eqref{eq:laplaceenvelope},~\eqref{eq:serviceunderflowprofile} are individual parameters. We add subscripts to clarify this where necessary.

In the following we evaluate the age of homogeneous parallel queues. Heterogeneous queues can be dealt with in the same way under the expense of additional notation.
%
%
\subsection{M$\mid$D$\mid$1 Queues}
\label{sec:md1}
For a Poisson packet arrival process with constant length packets $l > 0$ and mean inter-arrival time $w$, we have $\mathsf{P}[A(\tau,t)= k l] = e^{-(t-\tau)/w} ((t-\tau)/w)^k / k!$ and it follows that, e.g.,~\cite{fidler:netcalcguide},
\begin{equation}
\rho_A(\theta) = \frac{e^{\theta l} - 1}{\theta w} .
\label{eq:rhoapoisson}
\end{equation}

Further, from the Poisson distribution $\mathsf{P}[A(t,t+u) = 0] = e^{-u/w}$ and the lower arrival envelope in Def.~\ref{def:envelopes} has underflow profile
\begin{equation}
\underline{\varepsilon}_A(u) = e^{-u/w} .
\label{eq:underflowprofile}
\end{equation}

For constant rate service with rate $r > 0$ we insert~\eqref{eq:overflowprofile},~\eqref{eq:rhoapoisson},~\eqref{eq:underflowprofile} into Cor.~\ref{cor:constantrateservice} and obtain $\mathsf{P}[\Delta(t) > x] \le e^{-\theta (r x - l)} + e^{-(x - l/r)/w}$ for any $\theta > 0$ that satisfies $e^{\theta l} -1 \le \theta w r$. In a numerical evaluation, we optimize the free parameter $\theta$ to obtain the smallest upper bound.

In case of a parallel system with $k$ subsystems, the Poisson arrival process is split randomly and assigned to the subsystems $i \in [1,k]$ with probabilities $p_i = 1/k$. The result is independent Poisson processes with mean inter-arrival time $kw$. Hence, we can apply~\eqref{eq:ageindependent} to derive the tail distribution of the age of the parallel system.

\begin{figure}
\centering
\includegraphics[width=0.51\linewidth]{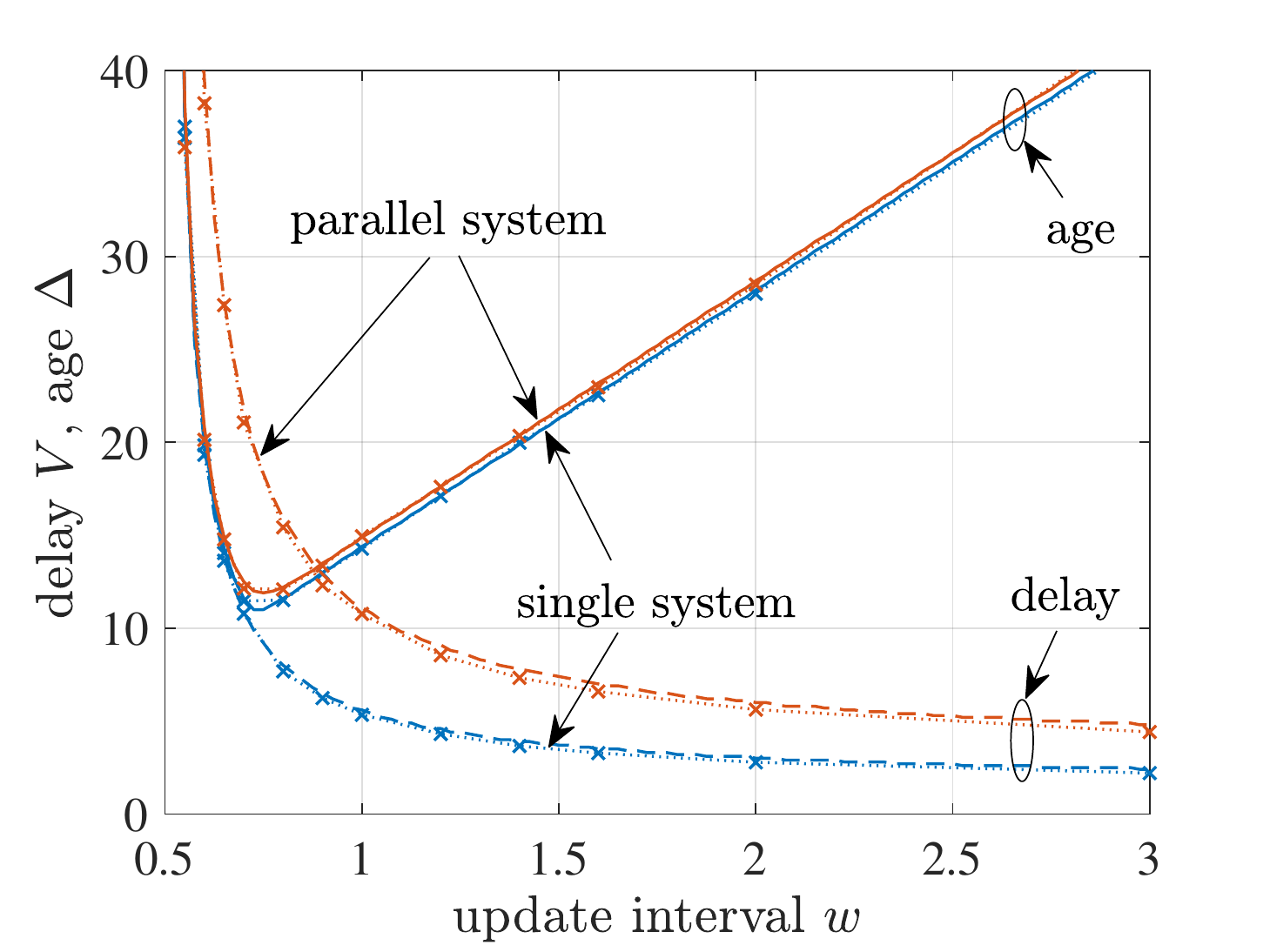}
\caption{Single M$\mid$D$\mid$1 queue with service rate $r=2$ versus two parallel M$\mid$D$\mid$1 queues each with rate $r=1$. Bounds with $\varepsilon=10^{-6}$ (age solid lines, delay dashed lines) compared to simulation results (dotted lines with markers).}
\label{fig:md1}
\end{figure}

We show numerical results for a single system with service rate $r=2$ and for a parallel system with two subsystems each with service rate $r=1$ in Fig.~\ref{fig:md1}. The packet size is $l=1$. Generally if not stated otherwise, tail bounds $\mathsf{P}[\Delta(t) > x] \le \varepsilon$ are computed for $\varepsilon = 10^{-6}$ and simulation results are obtained from $10^9$ samples. The virtual delay~\eqref{eq:virtualdelay} is depicted for reference.

In Fig.~\ref{fig:md1} we notice that age and delay are smaller in case of the M$\mid$D$\mid$1 compared to the M$\mid$M$\mid$1 queues. Further the minimal age of the M$\mid$D$\mid$1 queues is reached at a higher utilization and the bend of the age at its minimum is more pronounced, suggesting careful choice of the update interval. Apart from that, a similar behavior is observed for parallel M$\mid$D$\mid$1 queues as already seen in Fig.~\ref{fig:mm1_exact} for M$\mid$M$\mid$1 queues. The delay of the parallel system is worse than for the single system, which is a consequence of dividing the capacity, whereas both systems perform comparably with respect to the age. The effect that benefits the age in parallel systems but not the delay is that packets may overtake each other when scheduled randomly onto different subsystems. In case of the age, if packet $n+1$ departs earlier than packet $n$, it reduces the age and renders the late packet $n$ obsolete. This causes the age of the parallel M$\mid$D$\mid$1 queue to be almost on a par with the single M$\mid$D$\mid$1 queue with sum-equivalent service rate. A slight increase of the age of the parallel system is due to the increased packet service time.
%
%
\subsection{M$\mid$$\text{E}_l$$\mid$1 and M$\mid$M$\mid$1 Queues}
\label{sec:mm1}
We use the envelope of the Poisson arrival process from Sec.~\ref{sec:md1} and combine it with an envelope of a Poisson service process. For a Poisson service process with mean rate $r > 0$ it holds that $\mathsf{P}[S(\tau,t)= k] = e^{-r(t-\tau)} (r(t-\tau) )^k / k!$ and with the MGF of the Poisson process
\begin{equation}
\rho_S(\theta) = -\frac{r (e^{-\theta}-1)}{\theta} .
\label{eq:rhospoisson}
\end{equation}
Given packet arrivals with packet length $l$, it takes $k=l$ units of service of the Poisson service process to serve a packet. Thus, for Poisson arrivals with packet length $l$ we have an M$\mid$$\text{E}_l$$\mid$1 queue and for $l=1$ an M$\mid$M$\mid$1 queue.

For evaluation, we insert $\overline{\varepsilon}_A(b) = e^{-\theta_A b}$~\eqref{eq:overflowprofile}, $\varepsilon_S(b) = e^{-\theta_S b}$~\eqref{eq:serviceunderflowprofile}, $\underline{\varepsilon}_A(u) = e^{-u/w}$~\eqref{eq:underflowprofile}, and $\varepsilon_T(u) = e^{-\theta_T \rho_S(\theta_T) u}$ into Th.~\ref{th:aoi}, which holds for any $\theta_A, \theta_S, \theta_T > 0$ that satisfy $\rho_A(\theta_A) \le \rho_S(\theta_S)$ given in~\eqref{eq:rhoapoisson} and~\eqref{eq:rhospoisson}. The solution is obtained with Lem.~\ref{lem:convolution}. Finally, we optimize the free parameters $\theta$ and use~\eqref{eq:ageindependent} for parallel systems.

\begin{figure}
\hspace*{-0.5em}
\subfigure[M$\mid$M$\mid$1 queue, tail bounds]{
\includegraphics[width=0.51\linewidth]{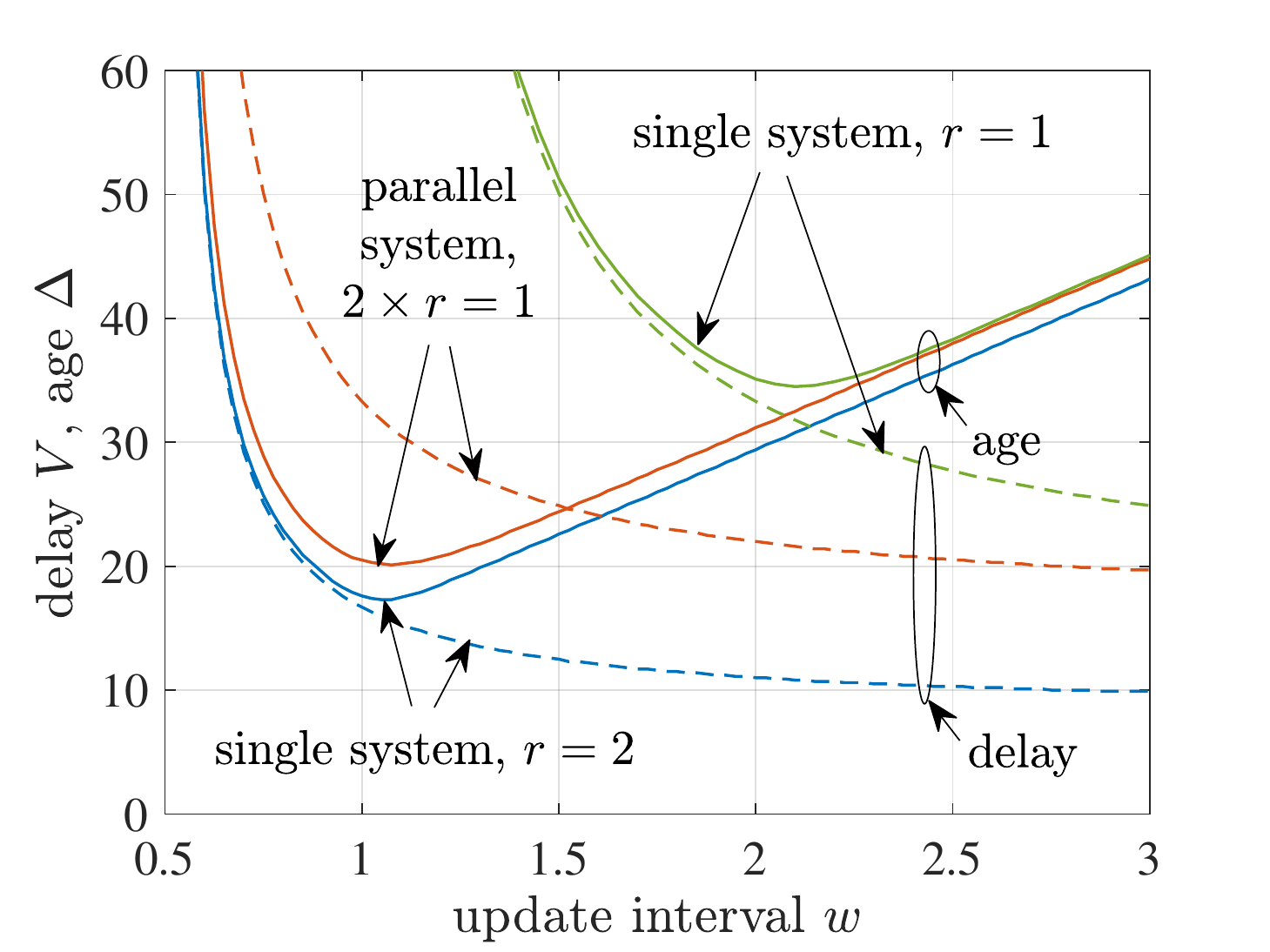}
\label{fig:mm1_bound}
}\hspace*{-1em}
\subfigure[M$\mid$M$\mid$1 queue, tail decay]{
\includegraphics[width=0.51\linewidth]{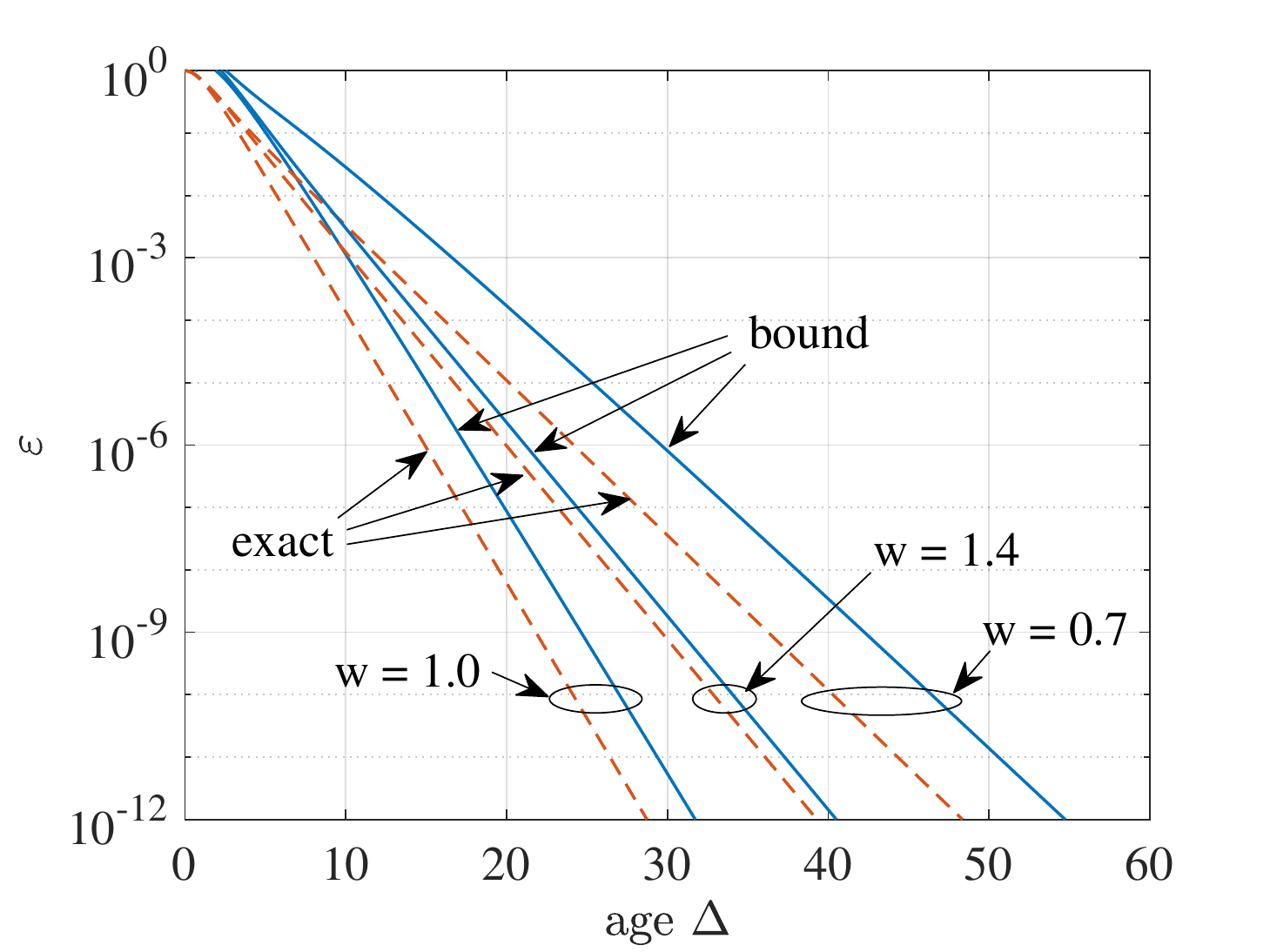}
\label{fig:mm1_tail}
}
\caption{(a) Tail bounds for $\varepsilon = 10^{-6}$ of single and parallel M$\mid$M$\mid$1 queues as in Fig.~\ref{fig:mm1_exact}. (b) Tail decay of a single M$\mid$M$\mid$1 queue with service rate $r=2$. Bounds compared to exact results. For update interval $w \approx 1$ the age is minimal.}
\label{fig:mm1_again}
\end{figure}
We showed results for parallel M$\mid$M$\mid$1 queues already in Fig.~\ref{fig:mm1_exact}, and here we only add numerical results that evaluate the accuracy of the bound. Fig.~\ref{fig:mm1_bound} reproduces bounds for the same systems as in Fig.~\ref{fig:mm1_exact}. Fig.~\ref{fig:mm1_tail} shows the tail bound compared to the exact tail decay of the M$\mid$M$\mid$1 with service rate $r=2$ queue. Compared to the results for the M$\mid$D$\mid$1 queue in Fig.~\ref{fig:md1}, it can be noticed that the bounds for the M$\mid$M$\mid$1 queue are slightly looser if $w$ is small. This is due to the use of Def.~\ref{def:envelopes} that specifies sample-path envelopes for the arrivals and for the service separately, see~\cite{rizk:statmuxing} for a discussion of the accuracy. We opted for the envelope-based approach since it maintains the modularity of the min-plus network calculus, which enables combining a variety of non-trivial arrival and service models with each other.
%
%
\subsection{D$\mid$$\text{E}_l$$\mid$1 and D$\mid$M$\mid$1 Queues}
\label{sec:dm1}
When we insert the parameters of the Poisson service process~\eqref{eq:serviceunderflowprofile} and~\eqref{eq:rhospoisson} into Cor.~\ref{cor:periodicarrivals} we have a D$\mid$$\text{E}_l$$\mid$1 queue and with packet length $l=1$ a D$\mid$M$\mid$1 queue. For deterministic arrivals Cor.~\ref{cor:periodicarrivals} takes the phase of the arrivals into account. This is important when considering parallel systems. We insert the time $t \ge 0$ that maximizes the tail bound given by Cor.~\ref{cor:periodicarrivals}. For a single D$\mid$M$\mid$1 queue we use a phase offset of $o=0$ and choose any $t$ that satisfies $(t-x) \bmod w = 0$.

In case of two (or similarly for more) parallel D$\mid$M$\mid$1 queues we use round robin scheduling so that the arrivals to the subsystems are also deterministic. Hence each queue sees an update interval of $2w$ and the second queue has a phase offset of $o=w$ compared to the first queue. Considering as before the time $t \ge 0$ that maximizes the tail bound, we can choose any $t$ that satisfies $(t-x) \bmod (2w) = 0$ for the first queue and hence $(t-x+w) \bmod (2w) = w$ since $o=w$ for the second queue, or vice versa. Any $t$ that does not satisfy this condition has a smaller tail bound. We use Cor.~\ref{cor:periodicarrivals} for each of the queues and since the service processes of the two parallel queues are independent~\eqref{eq:ageindependent} applies as before.

\begin{figure}
\hspace*{-0.5em}
\subfigure[D$\mid$M$\mid$1 queue, tail bounds]{
\includegraphics[width=0.51\linewidth]{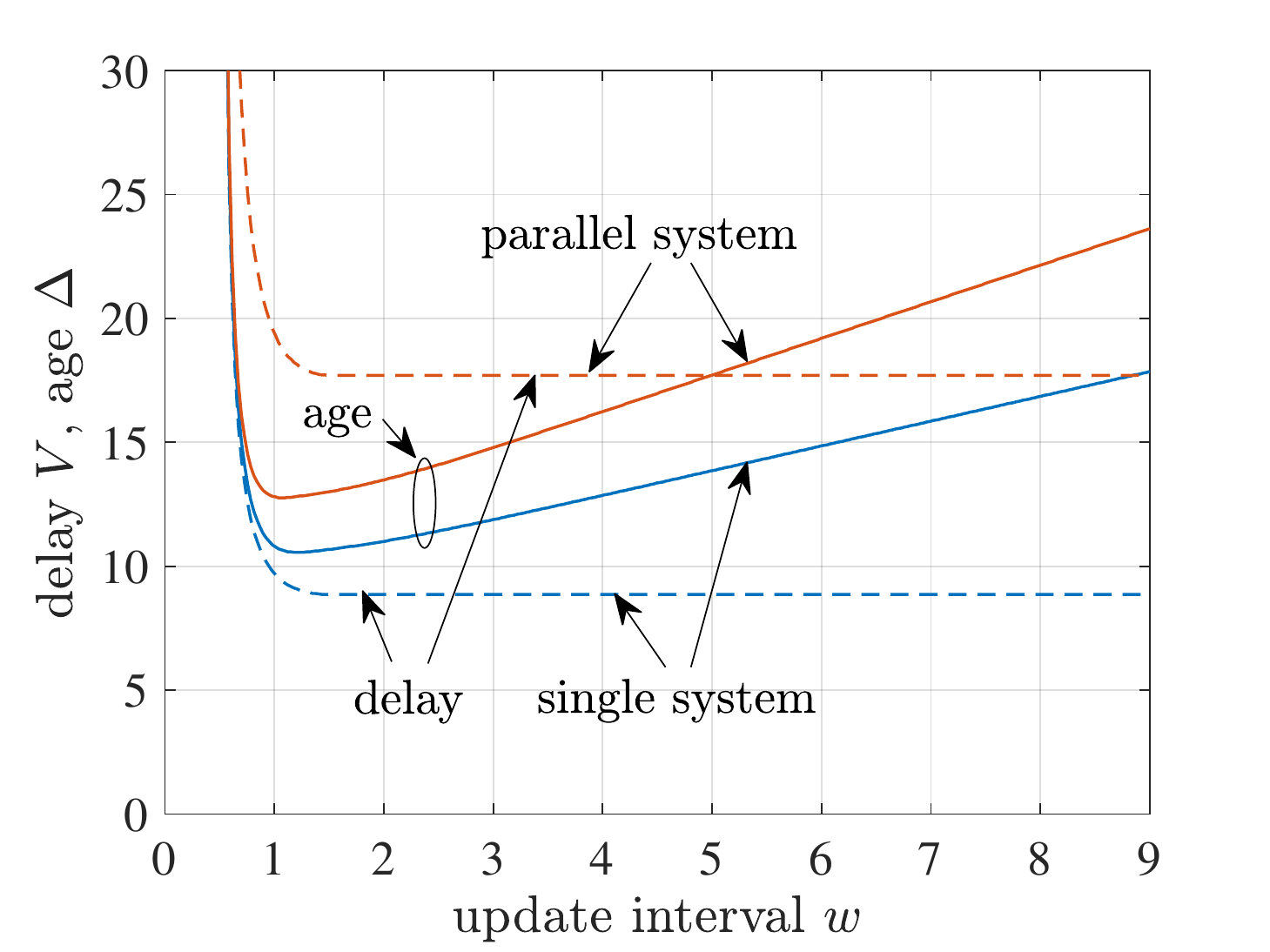}
\label{fig:dm1_bound}
}\hspace*{-1em}
\subfigure[D$\mid$M$\mid$1 queue, simulation]{
\includegraphics[width=0.51\linewidth]{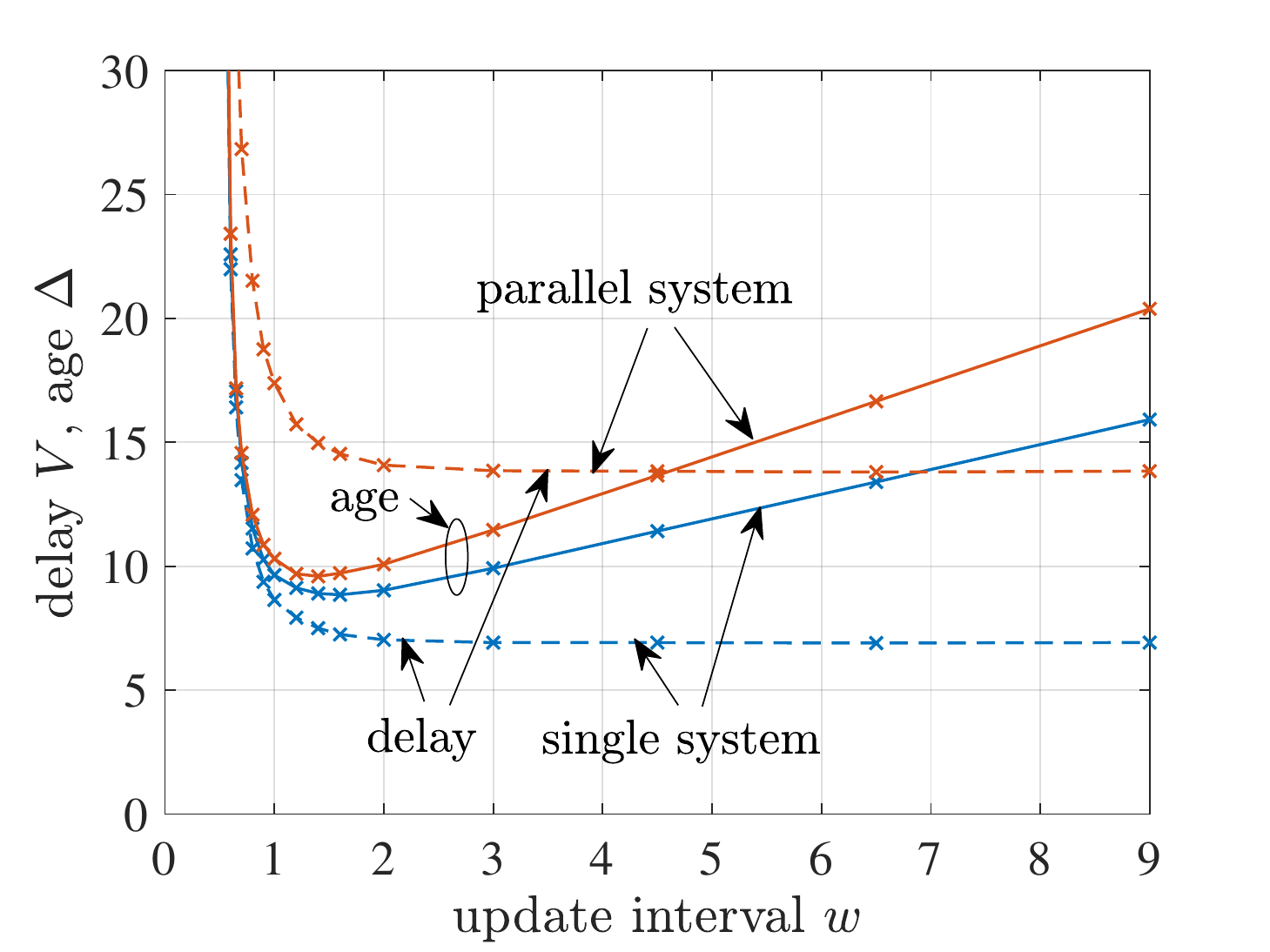}
\label{fig:dm1_sim}
}
\caption{Single D$\mid$M$\mid$1 queue with service rate $r=2$ versus two parallel D$\mid$M$\mid$1 queues each with rate $r=1$.}
\label{fig:dm1}
\end{figure}
In Fig.~\ref{fig:dm1} we show numerical results for a single D$\mid$M$\mid$1 queue with mean rate $r=2$ and for a system with two parallel D$\mid$M$\mid$1 queues each with rate $r=1$. The packet size is $l=1$. Since we consider $t$ that maximizes the bounds of age and delay, we show quantiles of the peak age and the packet delay (instead of age and virtual delay) obtained in simulations for comparison. The shape of the bounds matches that of the simulation results. A precondition for obtaining the correct slope of the age with increasing $w$ was to consider the phase of the arrivals at the parallel queues.

Regarding the age, we again notice that the parallel system is inferior, but unlike in case of the M$\mid$D$\mid$1 and the M$\mid$M$\mid$1 queue, the difference between the parallel and the single system is more pronounced and becomes larger with increasing $w$. Here, two effects occur. First, the increase in service time due to mean service rate $r=1$ for each of the servers in parallel queues in contrast to $r=2$ of the single system. This applies also in case of the M$\mid$D$\mid$1 and M$\mid$M$\mid$1 queues. Second, the contribution of the parallel D$\mid$M$\mid$1 queues via~\eqref{eq:ageindependent} diminishes with increasing $w$ due to the deterministic spacing caused by round robin scheduling of the periodic arrivals. It simply becomes unlikely that a delayed packet on one path can be compensated for by another packet on the other path, since that packet is already delayed by a phase offset of $w$.

Concluding, we find that in all cases, for M$\mid$D$\mid$1, M$\mid$M$\mid$1, and D$\mid$M$\mid$1 queues, the advantage that independent parallel systems take from~\eqref{eq:ageindependent} does not surpass the benefits of bundling the entire service rate into a single system. We also investigated more general parallel M$\mid$$\text{E}_l$$\mid$1 and D$\mid$$\text{E}_l$$\mid$1 queues. In essence, since the variability of Erlang-$l$ service is reduced if $l$ and $r$ are increased proportionally, the age gets smaller. In addition, the minimal age is achieved with a smaller update interval. Numerical as well as simulation results are shown in Fig.~\ref{fig:me21} in the appendix. The graphs do not exhibit a different behavior for parallel systems compared to a single system with sum-equivalent capacity than what we already observed for M$\mid$M$\mid$1 and D$\mid$M$\mid$1 queues.
%
%
\section{Parallel Markov Channels}
\label{sec:markovchannels}
For a more comprehensive investigation, we include Markov modulated service processes that have memory and hence may be considered less friendly. Specifically, in Sec.~\ref{sec:onoffchannels} we consider on-off channels, where the memory of the channel may cause long off periods during which the availability of a second channel can avoid excessive growth of the age. We consider various aspects of practical implementations including scheduling in Sec.~\ref{sec:scheduling} and periodic updates in Sec.~\ref{sec:periodicarrivals}. In Sec.~\ref{sec:heterogeneouschannels} we explore the use of heterogeneous channels with weighted splitting that is a relevant case in, e.g., 5G networks where a high rate but less reliable millimeter-wave channel may be combined with a lower rate but more reliable sub-6 GHz channel.

For analysis of these systems we first need to obtain a service envelope of a Markov modulated service process that conforms with Def.~\ref{def:envelopes}. For this, we employ a Martingale bound of the backlog derived in~\cite{poloczek:servicemartingales}. The backlog at time $t \ge 0$ is defined as $B(t) = A(t)-D(t)$ and with~\eqref{eq:serviceprocess} $B(t) \le \sup_{\tau \in [0,t]} \{A(\tau,t) - S(\tau,t)\}$. It follows by substitution of a rate function $A(\tau,t) = \rho (t-\tau)$ that
\begin{align}
\mathsf{P}[B(t) > b] \le & \mathsf{P}\biggl[\sup_{\tau \in [0,t]} \{A(\tau,t) - S(\tau,t)\} > b\biggr] \label{eq:backlog} \\
= & \mathsf{P}[\exists \tau \in [0,t]: \rho (t-\tau) - S(\tau,t) > b] \nonumber
\end{align}
takes the form of Def.~\ref{def:envelopes}, i.e., any backlog bound provides a valid service envelope. We note that the non-negativity condition of the service envelope in Def.~\ref{def:envelopes} is satisfied trivially since $S(\tau,t)$ is non-negative. Further, for stationary processes it is known from~\cite[Lem. 9.1.4]{chang:performanceguarantees} that~\eqref{eq:backlog} is stochastically increasing with $t$. It follows that stationary upper bounds of~\eqref{eq:backlog} for $t \rightarrow \infty$ satisfy Def.~\ref{def:envelopes} for all $t \ge 0$. Hence, we can invoke the backlog bound~\cite[Th. 9]{poloczek:servicemartingales} with $A(\tau,t) = \rho_S  (t-\tau)$ to obtain the envelope parameters.

The service $S(\tau,t)$ is modulated by a discrete-time Markov chain with $n$ states and transition matrix $P$, that has elements $p_{ij} = \mathsf{P}[y_{t+1} = j \mid y_t = i]$ where $y_t$ denotes the state at time $t \ge 0$. The diagonal matrix $R = \text{diag} \bigl(e^{\theta r_1}, e^{\theta r_2}, \dots, e^{\theta r_n}\bigr)$ specifies the service rate $r_i$ in state $i$. It follows that the service envelope in Def.~\ref{def:envelopes} is satisfied with parameters~\cite[Ex. 8]{poloczek:servicemartingales}
\begin{equation}
\rho_S(\theta) = -\frac{\ln \text{sp}(P R(-\theta))}{\theta}
\label{eq:rhosmarkov}
\end{equation}
and
\begin{equation}
\varepsilon_S(b) = \frac{\mathsf{E}[h(y_0)]}{\min_{i:\rho_S(\theta)>r_i}\{h(i)\}} e^{-\theta b}
\label{eq:serviceunderflowprofilemarkov}
\end{equation}
where $\text{sp}$ denotes the spectral radius, $h$ is a corresponding right eigenvector, and $h(i)$ is the $i$-th component of $h$.
%
%
\subsection{On-Off Channels}
\label{sec:onoffchannels}
For numerical investigation we use the basic model of an on-off channel. In off state, i.e., state 1, the rate of the channel is $r_1 = 0$ and in on state $r_2 = c$. The steady-state probabilities are $p_{\text{on}} = p_{12}/(p_{12} + p_{21})$ and $p_{\text{off}} = 1-p_{\text{on}}$. The average service rate is $r=p_{\text{on}} c$ and the burstiness of the channel can be characterized by $\beta = 1/p_{12} + 1/p_{21}$ that is the mean time to change state twice. For a channel without memory $p_{11} = p_{21} = p_{\text{off}}$ and $p_{22} = p_{12} = p_{\text{on}}$ so that $\beta_0 = 1/(p_{\text{on}} p_{\text{off}})$. For an explicit solution of~\eqref{eq:rhosmarkov} for the on-off channel see~\cite[Eq. 7.18]{chang:performanceguarantees}. Statistical bounds of the age follow in the same way as for the M$\mid$E$_l$$\mid$1 queue in Sec.~\ref{sec:mm1}: we insert the parameters of the Poisson arrivals~\eqref{eq:overflowprofile},~\eqref{eq:rhoapoisson}, and~\eqref{eq:underflowprofile} and the parameters of the on-off service process~\eqref{eq:rhosmarkov} and~\eqref{eq:serviceunderflowprofilemarkov} into Th.~\ref{th:aoi}.

\begin{figure}
\hspace*{-0.5em}
\subfigure[$\beta = \beta_0$]{
\includegraphics[width=0.51\linewidth]{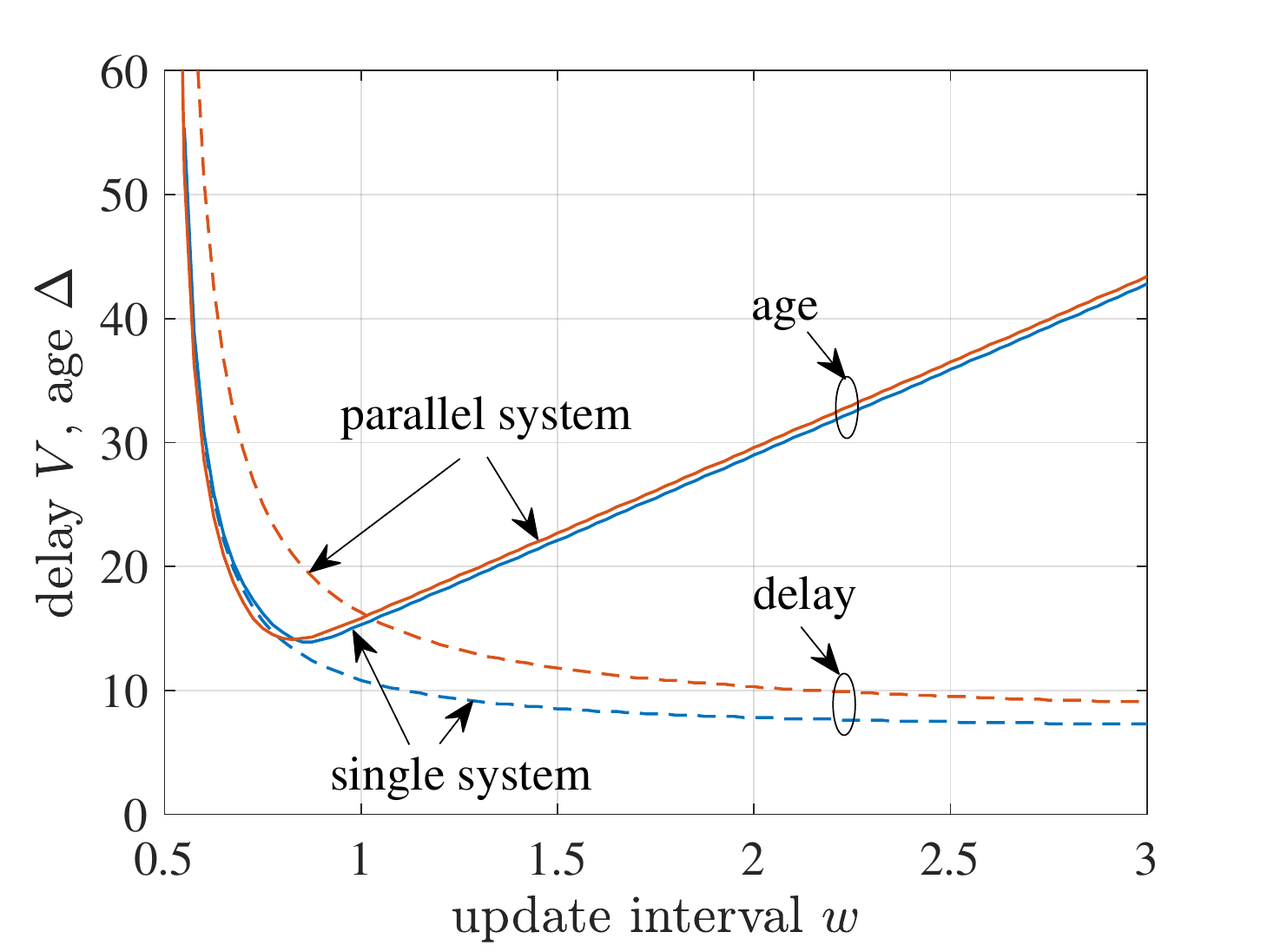}
\label{fig:onoffbeta1}
}\hspace*{-1em}
\subfigure[$\beta = 3 \beta_0$]{
\includegraphics[width=0.51\linewidth]{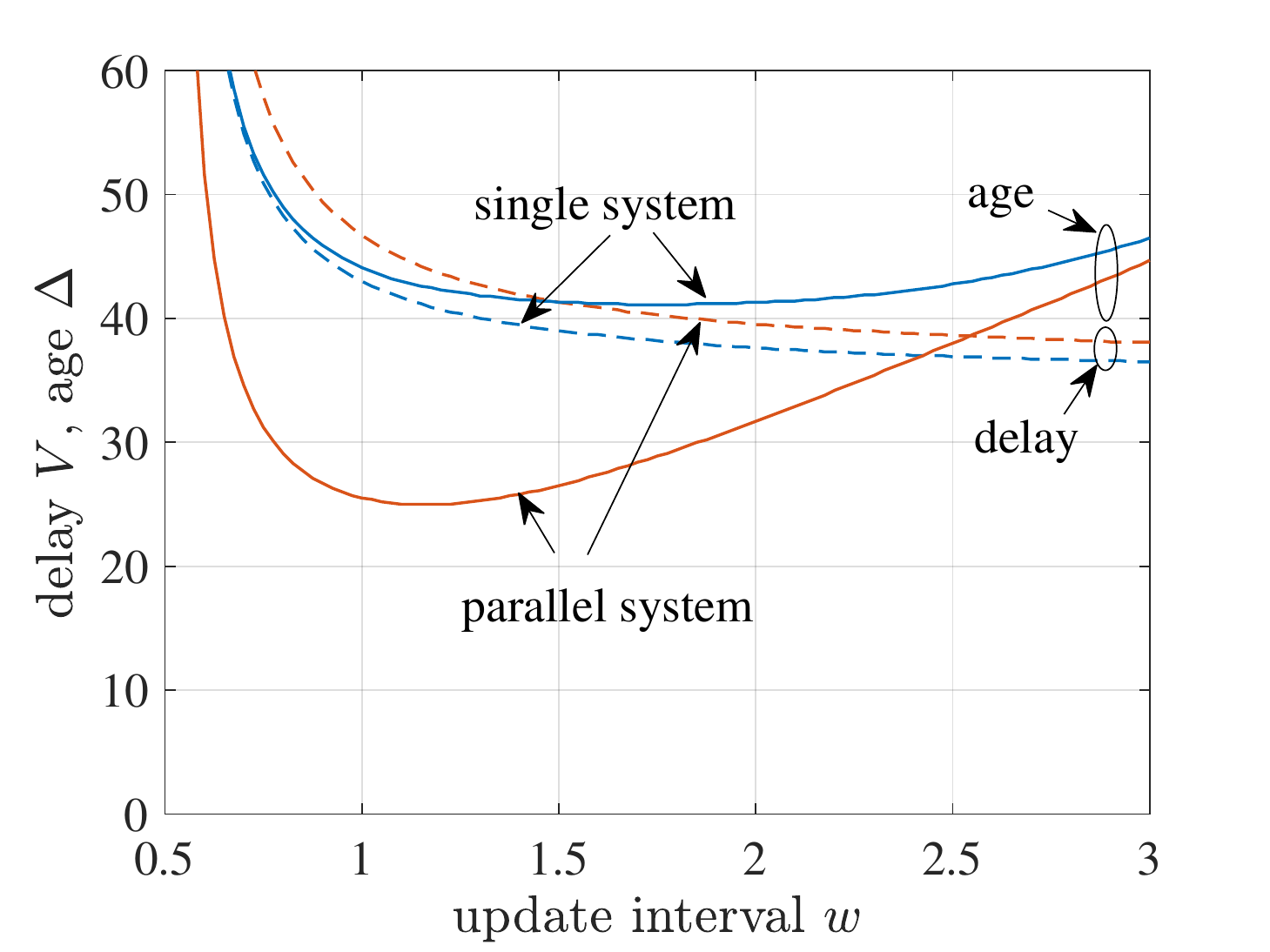}
\label{fig:onoffbeta3}
}\\\hspace*{-0.5em}
\subfigure[$\beta = 2 \beta_0$]{
\includegraphics[width=0.51\linewidth]{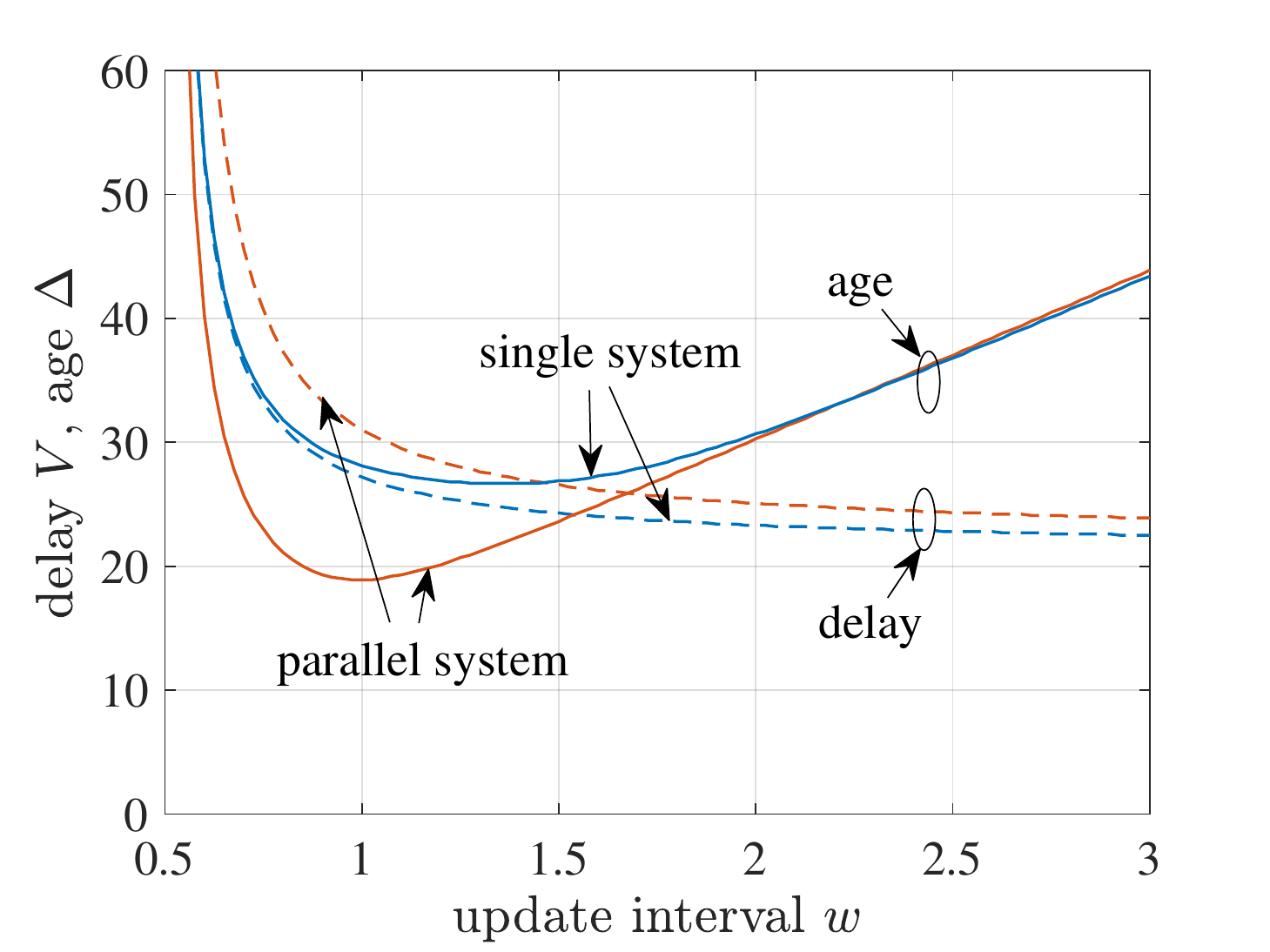}
\label{fig:onoffbeta2}
}\hspace*{-1em}
\subfigure[$\beta = 2 \beta_0$, simulation]{
\includegraphics[width=0.51\linewidth]{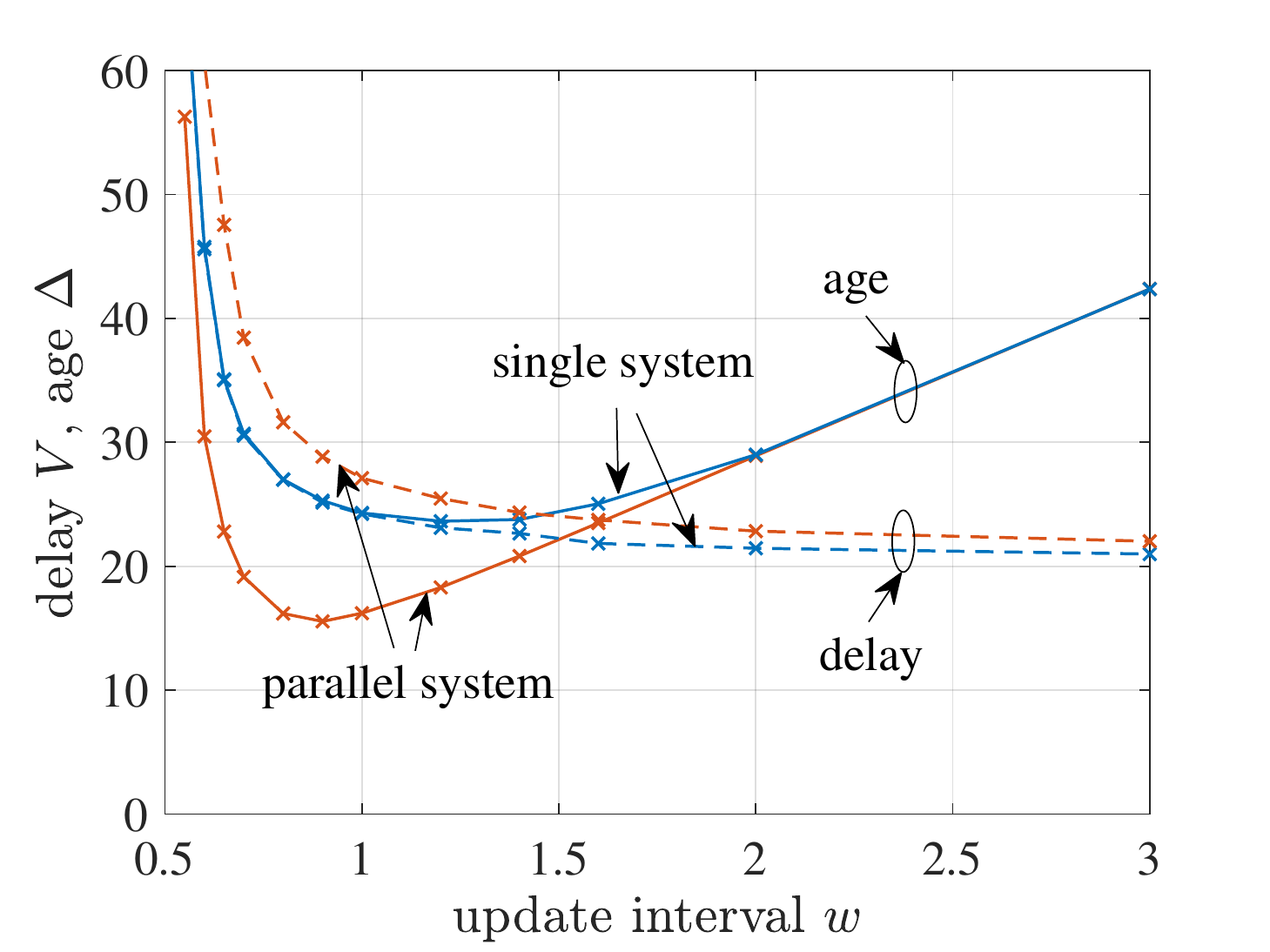}
\label{fig:onoffbeta2sim}
}
\caption{Poisson arrivals at a single Markov on-off channel with mean rate $r=2$ compared to a parallel system with two independent on-off channels each with mean rate $r=1$. The burstiness parameter of the channels $\beta$ has a large effect on the age and causes a fundamental change: the parallel system outperforms the sum-equivalent single system.}
\label{fig:onoff}
\end{figure}
In Fig.~\ref{fig:onoff} we depict the age and the virtual delay of a single on-off channel with mean rate $r=2$ compared to a parallel system with two independent on-off channels each with mean rate $r=1$. The remaining parameters of all channels are identical, $p_{\text{on}}=0.9$ and the channel burstiness $\beta$ is given with respect to the case without memory $\beta_0 = 11,\overline{1}$. The arrivals are Poisson with mean rate $1/w$ and packet length $l=1$. For $\beta = \beta_0$ in Fig.~\ref{fig:onoffbeta1} we notice that the age of the single and the parallel system is more or less on a par. In contrast, Figs.~\ref{fig:onoffbeta3} and~\ref{fig:onoffbeta2} show a fundamentally different behavior: if the memory of the channel increases, the parallel system achieves a significant advantage. In this case, it appears that the parallel system benefits largely from the independence of the channels whereas the single system takes less advantage of the cumulated service rate.
In Fig.~\ref{fig:onoffbeta2sim} we present the results from system simulation for $\beta = 2 \beta_0$ which confirms the trends provided by the tail bounds.

\begin{figure}
\hspace*{-0.5em}
\subfigure[Impact of the service rate]{
\includegraphics[width=0.51\linewidth]{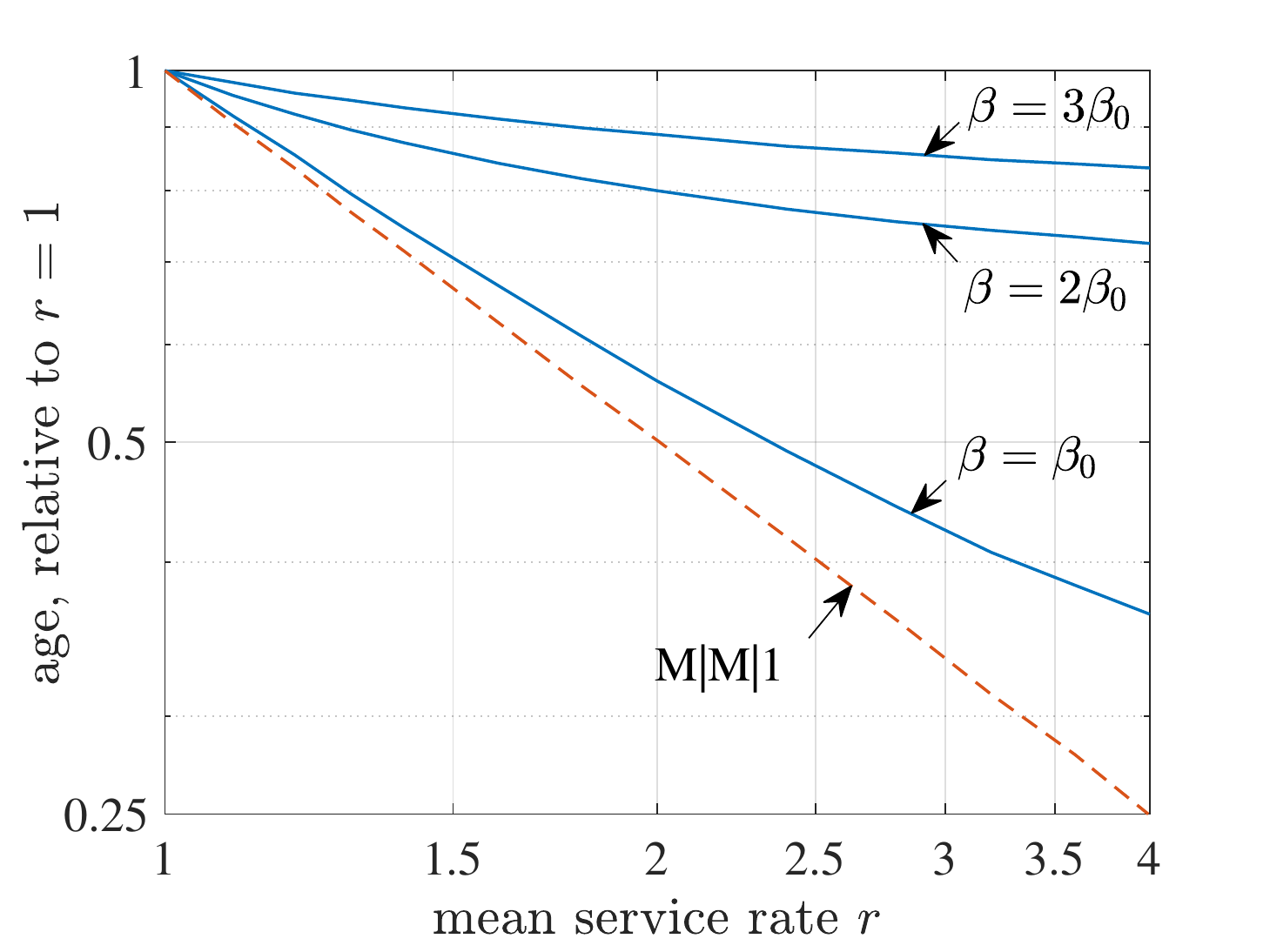}
\label{fig:onoff_capacity}
}\hspace*{-1em}
\subfigure[Tail decay]{
\includegraphics[width=0.51\linewidth]{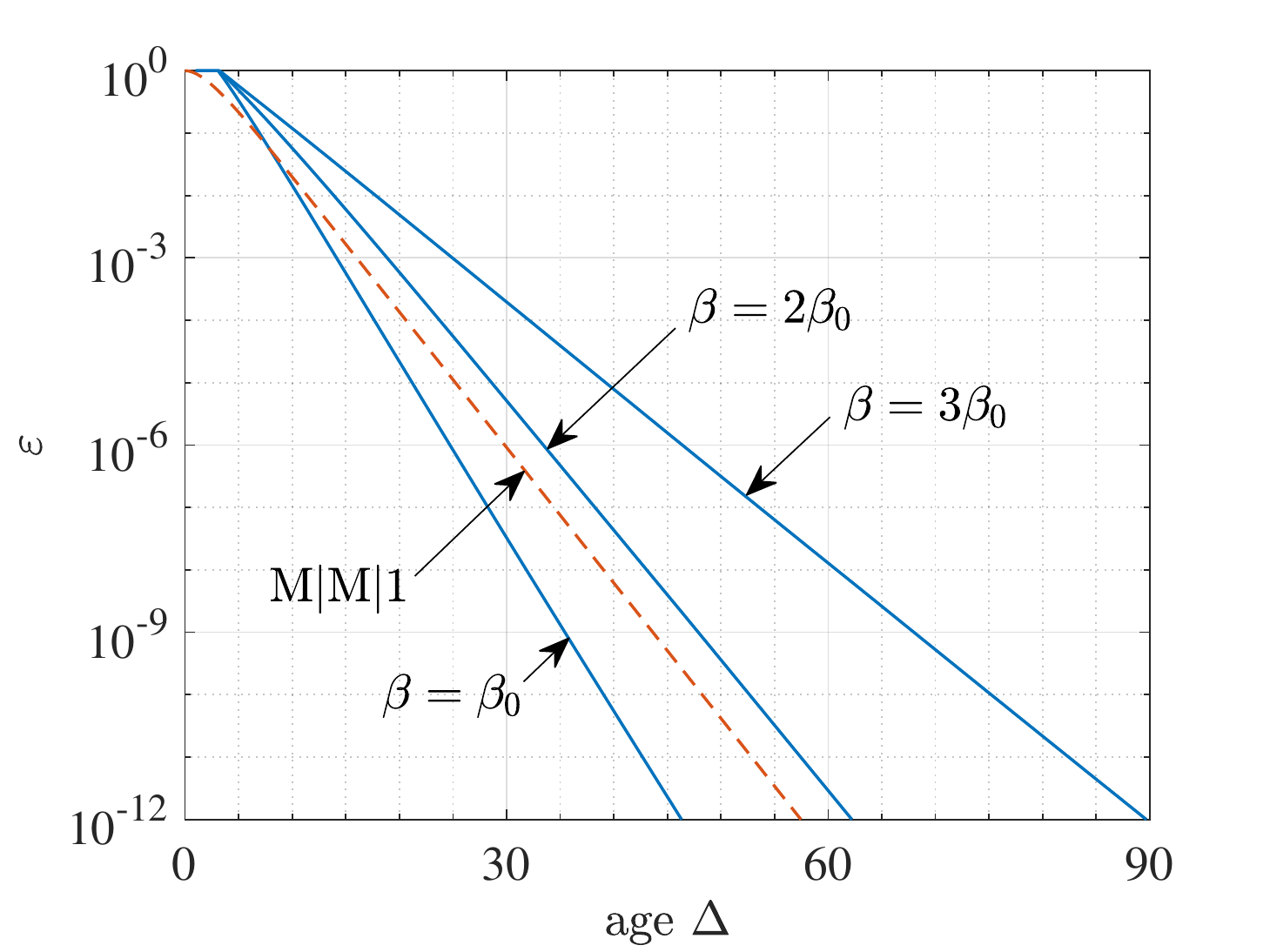}
\label{fig:onoff_tail}
}
\caption{Poisson arrivals at a single Markov on-off channel compared to the M$\mid$M$\mid$1 queue. (a) Age at $\varepsilon = 10^{-6}$ for increasing mean service rate $r$, relative to the age at $r=1$. (b) Tail decay of the age for $r=1$.}
\label{fig:onoff_tradeoff}
\end{figure}

To illustrate how this change due to memory of the channel comes about, Fig.~\ref{fig:onoff_tradeoff} contrasts two effects. We show age bounds for single on-off channels with $p_{\text{on}} = 0.9$. For reference the age of the M$\mid$M$\mid$1 queue~\cite{inoue:aoisingleserverqueues} is included. The update interval $w$ that minimizes the age at $\varepsilon = 10^{-6}$ is used. Fig.~\ref{fig:onoff_capacity} shows the effect if the service rate $r$ is increased. The age is depicted relative to the age at $r=1$ (the absolute age at $r=1$ can be read from Fig.~\ref{fig:onoff_tail}). We notice that the age of the M$\mid$M$\mid$1 queue is halved if $r$ is doubled. This determines the gain that a single system takes from the sum-equivalent service rate of a parallel system. For the case of an on-off channel without memory, $\beta = \beta_0$, the gain of extra service rate is slightly smaller and diminishes quickly as the channel burstiness $\beta$ increases. Intuitively, a higher service rate in on state -- even if it is much higher -- is of little use if the age is dominated by long off periods of the channel.

In contrast, the benefit of independent parallel channels comes from~\eqref{eq:ageindependent} and it is determined by the tail decay of the age that is depicted in Fig.~\ref{fig:onoff_tail} for $r=1$. In all cases the tail decay is exponential. This implies that the age is generally halved (slightly less due to a small offset at the origin) if we double the degree of parallelism. For an example, consider the age of the on-off channel for $\beta = 2\beta_0$. At $\varepsilon = 10^{-12}$ the age is $\Delta \approx 62$, using~\eqref{eq:ageindependent} we have $\varepsilon = (10^{-6})^2$ for two homogeneous parallel channels and $\Delta \approx 33$ and for four channels $\varepsilon = (10^{-3})^4$ and $\Delta \approx 18$. Concluding, given additional capacity, Fig.~\ref{fig:onoff_tradeoff} answers how this capacity can be used most effectively, whether on a single channel or on additional parallel channels.
%
%
\subsection{Scheduling Policies}
\label{sec:scheduling}
So far we used random splitting of Poisson arrival processes. This achieves independent Poisson arrival processes for the individual subsystems and ensures applicability of~\eqref{eq:ageindependent}. A natural question that arises is whether other scheduling policies than random may improve the age of parallel systems. We evaluate round robin scheduling and join the shortest queue scheduling. In case of round robin scheduling with $k$ subsystems the Poisson arrival stream is divided into $k$ streams that each have Erlang-$k$ inter-arrival times. The individual arrival streams are, however, not independent so that the assumption of~\eqref{eq:ageindependent} is not satisfied. Join the shortest queue scheduling requires additional feedback about the size of the queues from the subsystems. Using this feedback for scheduling, the arrivals at the subsystems will depend on the subsystems' service processes (the service in the past and in case of a channel with memory also the service in the future). This makes a fundamental difference for the analysis and we use simulation for further numerical evaluation.

\begin{figure}
\hspace*{-0.5em}
\subfigure[M$\mid$M$\mid$1 queue]{
\includegraphics[width=0.51\linewidth]{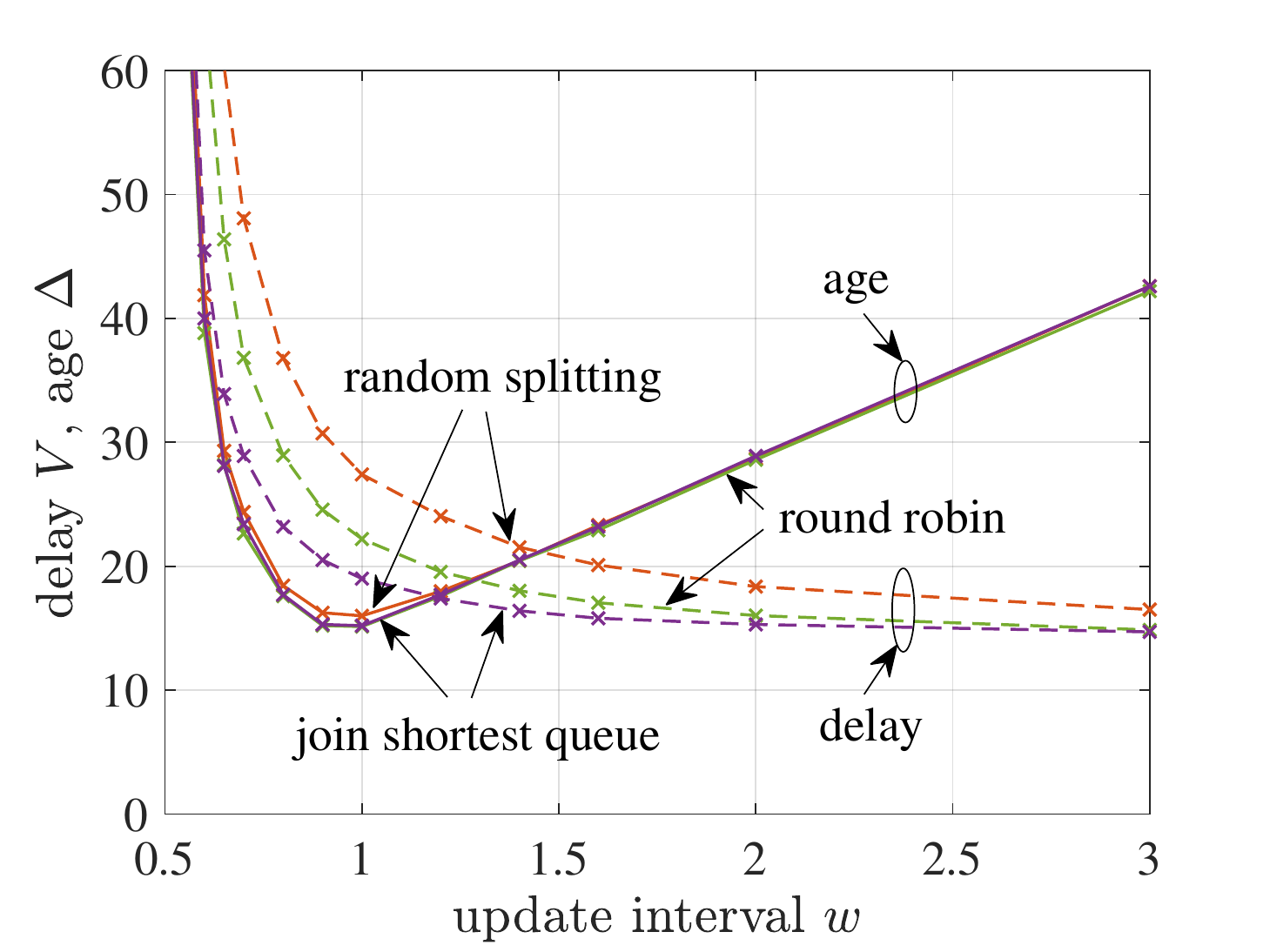}
\label{fig:mm1simscheduling}
}\hspace*{-1em}
\subfigure[On-off channel, $\beta = 2 \beta_0$]{
\includegraphics[width=0.51\linewidth]{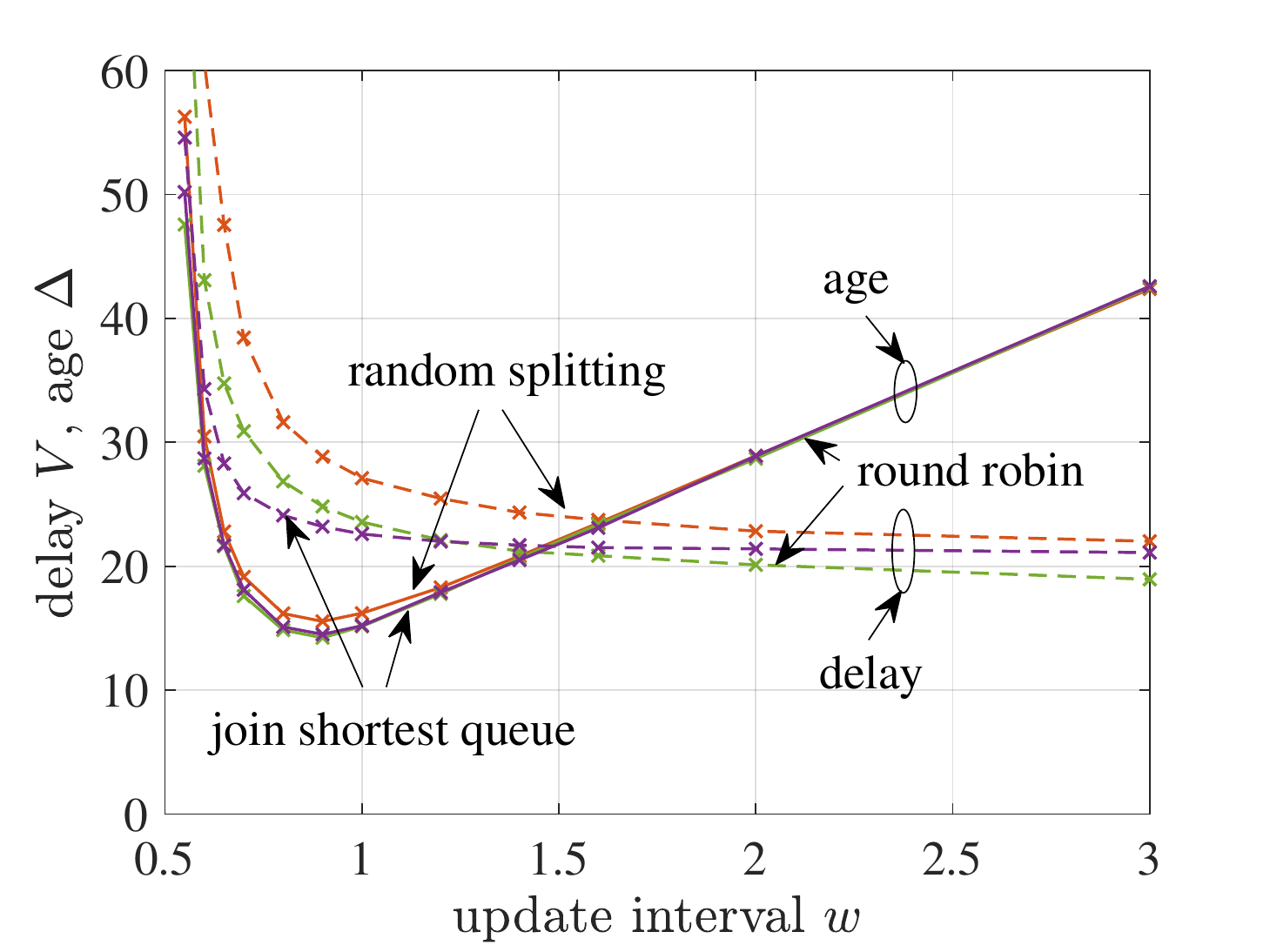}
\label{fig:onoffbeta2simscheduling}
}
\caption{Parallel systems with Poisson arrivals and random splitting compared to round robin and join the shortest queue scheduling. Scheduling is effective in reducing delays but has little impact on the age. The age achieved by round robin and by join the shortest queue scheduling is almost identical and close to random splitting.}
\label{fig:scheduling}
\end{figure}

In Fig.~\ref{fig:scheduling} we compare the impact of the scheduling policy for the M$\mid$M$\mid$1 queue as in Fig.~\ref{fig:mm1_exact}, and for the on-off channel with burstiness $\beta = 2 \beta_0$ as in Fig.~\ref{fig:onoffbeta2sim}. We do not include a graph for the M$\mid$D$\mid$1 queue as it does not show any other effect. Compared to random splitting, round robin scheduling achieves noticeably smaller delay quantiles. This is a consequence of the more even distribution of the load. For small $w$, i.e., high utilization join the shortest queue improves the delay even more. For the age, we find that the effect of the scheduling policy is minor, however. In case of the M$\mid$M$\mid$1 queue the parallel system did not improve the age compared to the single system with sum-equivalent capacity, see Fig.~\ref{fig:mm1_exact}, and replacing the scheduling policy does not change this. In case of the bursty on-off channel the parallel system achieves a large improvement of the age over the single system, see Fig.~\ref{fig:onoffbeta2sim}. This applies for all scheduling policies including random splitting and the differences between the scheduling policies are again small.

An explanation for the good performance of random splitting is that in the event of congestion, an imbalance of the load is not that bad for the age since packets on the less congested path may overtake packets on the congested path and improve the age. This does not help with the delay, though. While the choice of the scheduling policy can help reduce the delay quantiles significantly, we conclude from our experiments that the overhead of scheduling, e.g., feedback in case of join the shortest queue, pays off very little for the age.
%
%
\subsection{Periodic Arrivals}
\label{sec:periodicarrivals}
It is well-known that the age of periodic arrivals outperforms Poisson arrivals, e.g., compare the age of the single systems in Figs.~\ref{fig:dm1} and~\ref{fig:mm1_exact}, and it is interesting to see how this plays out in case of parallel on-off channels. For periodic arrivals and round robin scheduling we use Cor.~\ref{cor:periodicarrivals}, where we insert the parameters of the on-off channel~\eqref{eq:rhosmarkov} and~\eqref{eq:serviceunderflowprofilemarkov}. As before in Sec.~\ref{sec:dm1} for parallel D$\mid$M$\mid$1 systems, we pay attention to the phase of the arrivals, and we consider the time $t$ that maximizes the bounds. Correspondingly, we show simulation results of peak age and packet delay for comparison.

\begin{figure}
\hspace*{-0.5em}
\subfigure[$\beta = \beta_0$]{
\includegraphics[width=0.51\linewidth]{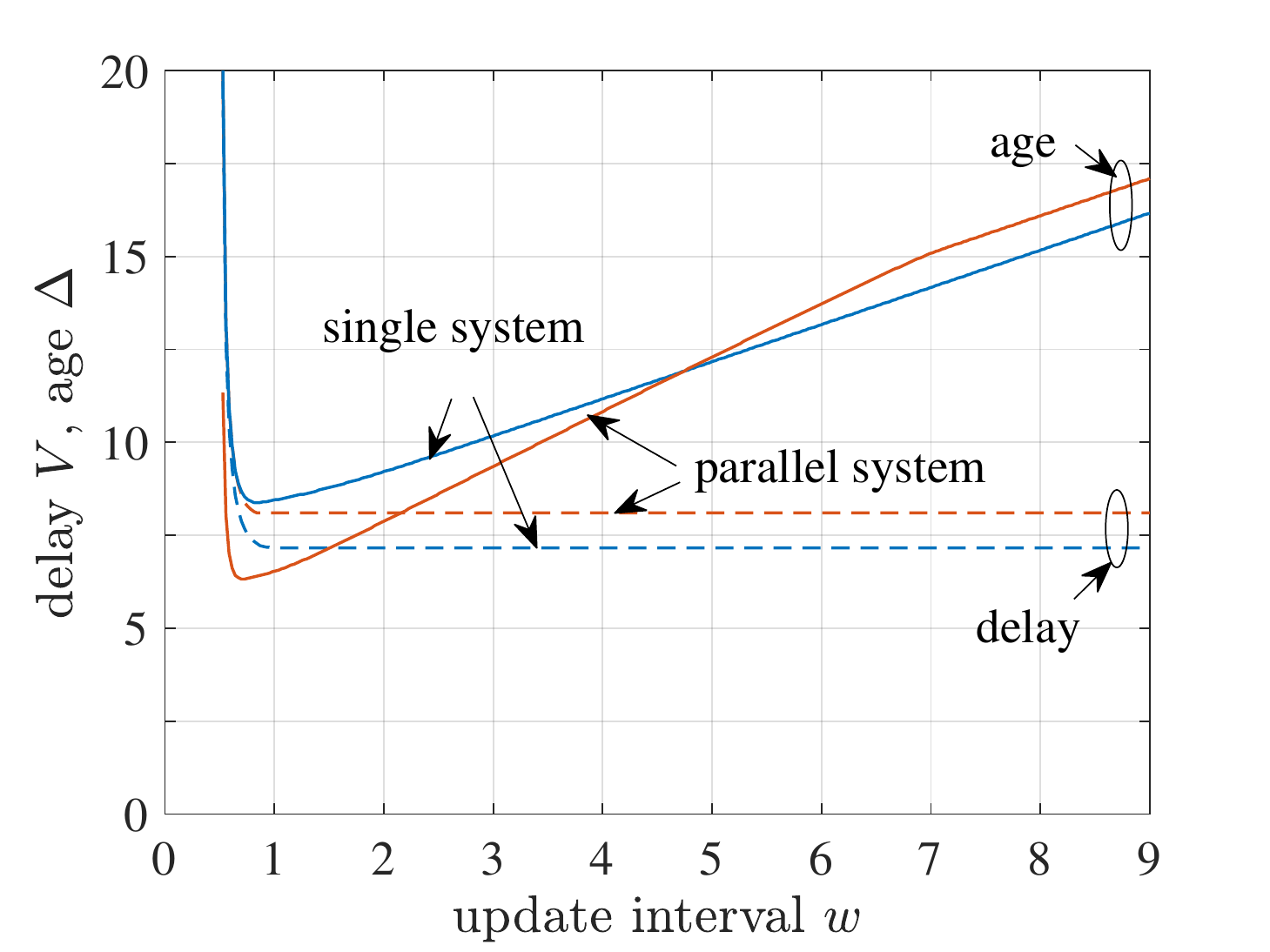}
\label{fig:onoffbeta1periodic}
}\hspace*{-1em}
\subfigure[$\beta = \beta_0$, simulation]{
\includegraphics[width=0.51\linewidth]{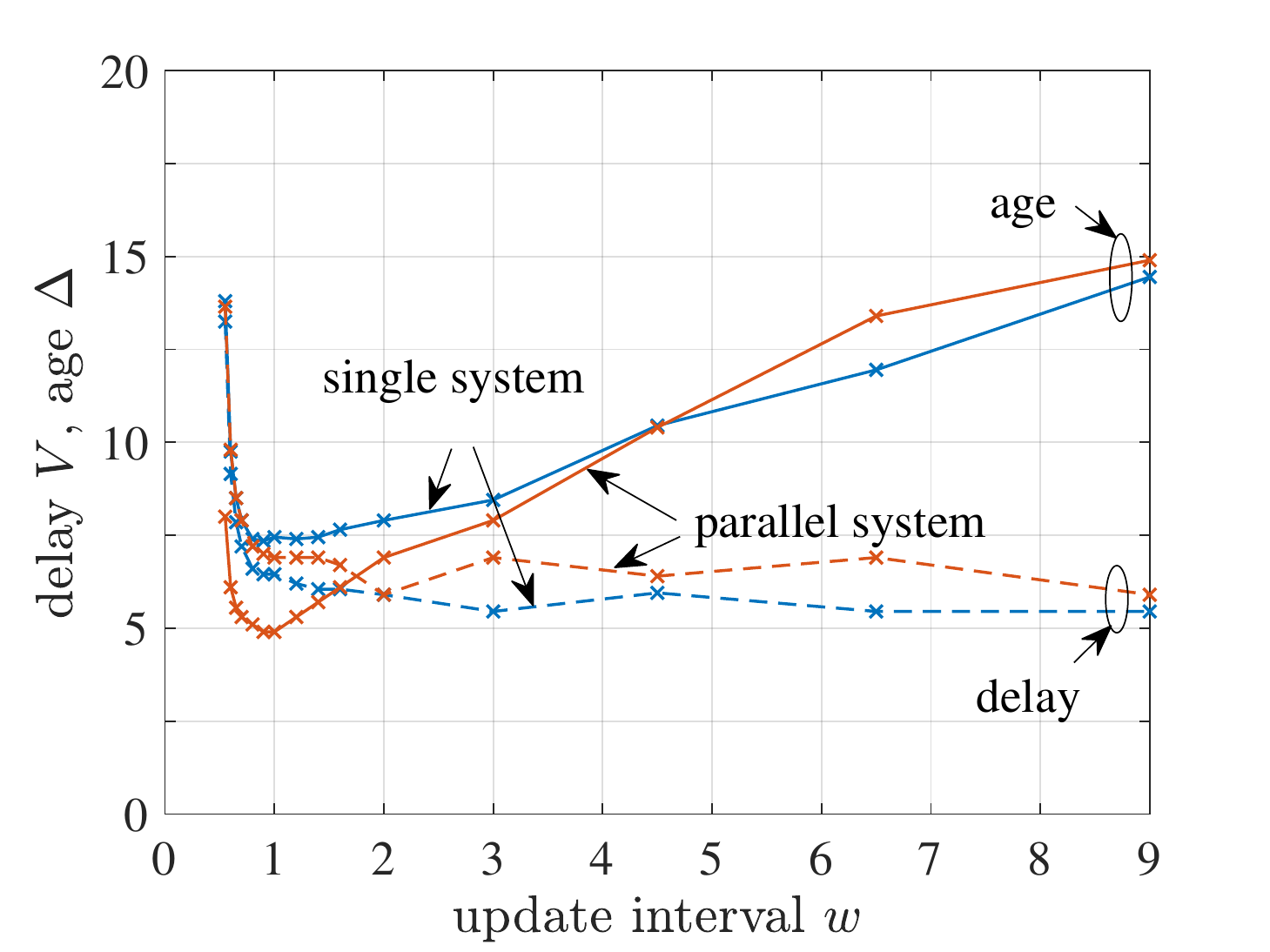}
\label{fig:onoffbeta1periodicsim}
}\\\hspace*{-0.5em}
\subfigure[$\beta = 2 \beta_0$]{
\includegraphics[width=0.51\linewidth]{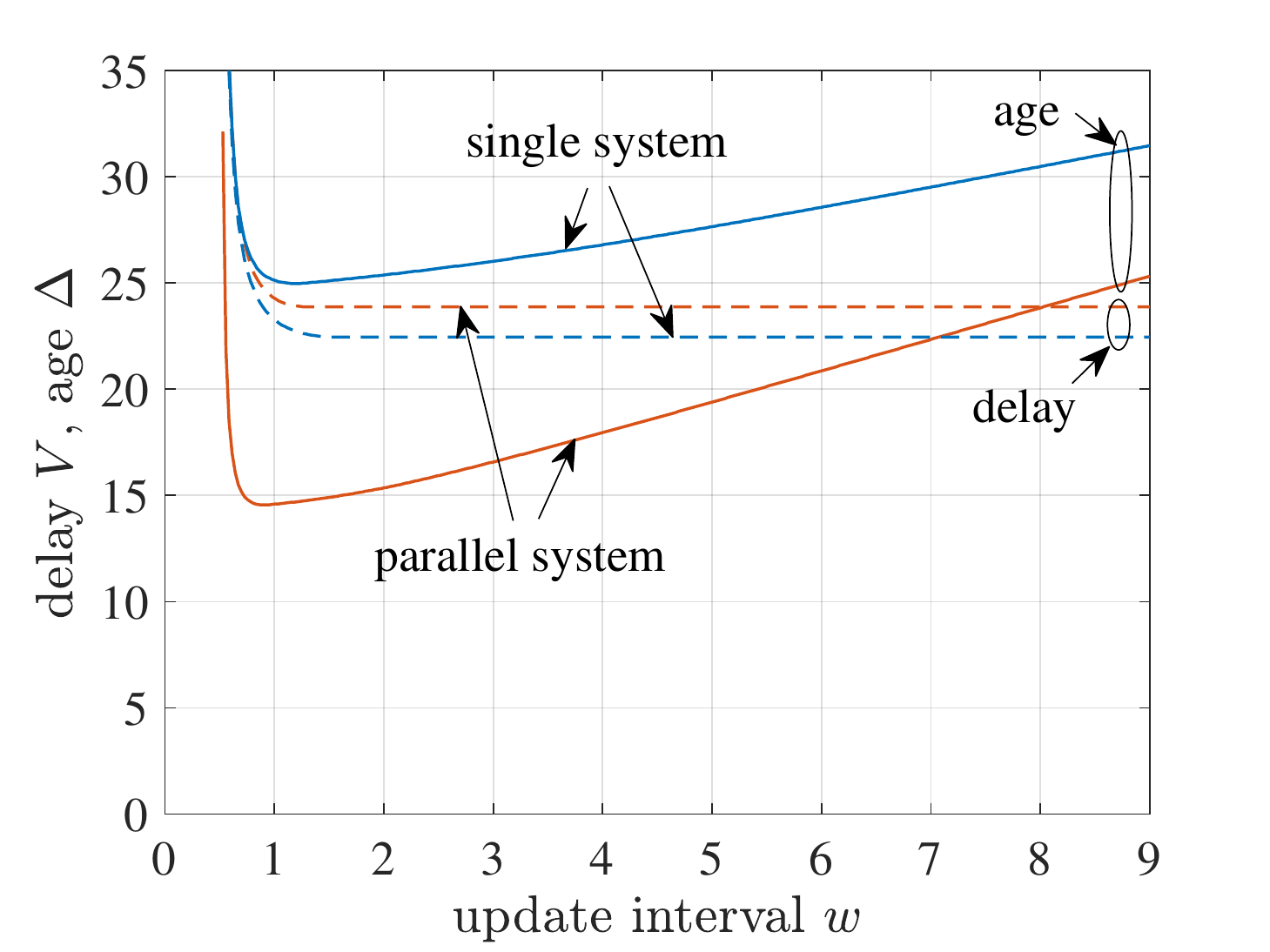}
\label{fig:onoffbeta2periodic}
}\hspace*{-1em}
\subfigure[$\beta = 2 \beta_0$, simulation]{
\includegraphics[width=0.51\linewidth]{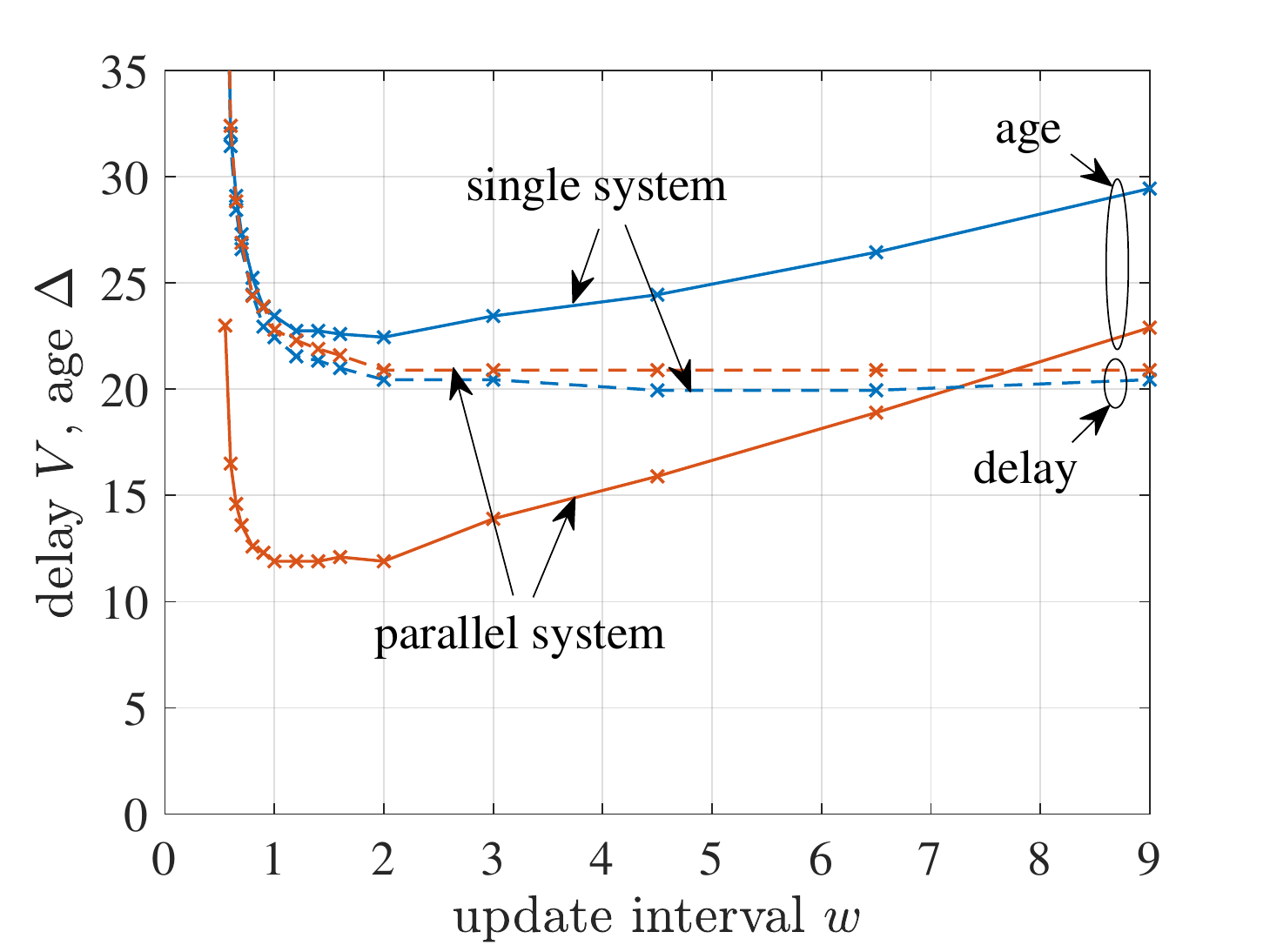}
\label{fig:onoffbeta2periodicsim}
}
\caption{Periodic arrivals at a single Markov on-off channel with mean rate $r=2$ compared to a parallel system with two independent on-off channels each with mean rate $r=1$ and round robin scheduling.}
\label{fig:periodic}
\end{figure}

Numerical and simulation results for on-off channels with the same parameters as in Fig.~\ref{fig:onoff} but now for periodic arrivals and round robin scheduling are shown in Fig.~\ref{fig:periodic}. For a single bursty channel with $\beta = 2 \beta_0$ we observe in Fig.~\ref{fig:onoffbeta2periodic} that periodic arrivals generally achieve a somewhat smaller minimal age than Poisson arrivals in Fig.~\ref{fig:onoffbeta2}. A much more significant improvement of the age is obtained, however, if we compare the single system to the parallel system, for periodic and for Poisson arrivals alike.

Interestingly, for the case of an on-off channel without memory, $\beta = \beta_0$, and periodic arrivals Fig.~\ref{fig:onoffbeta1periodic} shows a different behavior than Fig.~\ref{fig:onoffbeta1} for Poisson arrivals. The minimal age that is achieved by the parallel system is smaller than that of the single system. Corresponding simulation results in Fig.~\ref{fig:onoffbeta1periodicsim} confirm this effect. As before, the simulation uses $10^9$ packet samples but since the age values are quite small, it becomes increasingly difficult to obtain accurate $(1-10^{-6})$-quantiles.

The advantage of the parallel system in Fig.~\ref{fig:onoffbeta1periodic} vanishes, however, with increasing $w$ and starting at $w = 5$ the age of the parallel system becomes worse than that of the single system. The reason for this is that the independence of the parallel channels helps little, if at all, as $w$ increases. For illustration, assume packet $n-1$ is available at the monitor, packet $n$ sees an excessively long delay on channel 1, whereas packet $n+1$ on channel 2 is not delayed. Packet $n+1$ arrives earliest at time $T_D(n+1) = T_A(n-1) + 2w + l/c$ when the age of packet $n-1$ has reached at least $2w+l/c$. For any age threshold $\mathsf{P}[\Delta(t) > x]$ where $x < 2w+l/c$ this is too late. In Figs.~\ref{fig:onoffbeta1periodic} and~\ref{fig:onoffbeta1periodicsim} we see how this effect makes the age of the parallel system inferior at $w=5$ on the abscissa and above $2w+l/c \approx 11$ on the ordinate axis.
%
%
\subsection{Heterogeneous Channels with Weighted Splitting}
\label{sec:heterogeneouschannels}
Heterogeneous channels arise in many relevant cases, such as in 5G New Radio networks, where channels in frequency range FR1 at sub-6 GHz and FR2 millimeter-wave may be used in parallel, or in multi-technology wireless access using, e.g., cellular and Wifi. We use the Markov on-off model to characterize these wireless channels and evaluate if and how channels with heterogeneous parameters may benefit the age. We consider a parallel system with two channels: channel 1 has low mean service rate $r = 1$ with few outages $p_{\text{on}} = 0.95$ and $\beta = \beta_0$, whereas channel 2 has high mean service rate $r=4$ with more frequent outages $p_{\text{on}} = 0.9$, and memory $\beta = 2 \beta_0$. The arrivals are Poisson and are split randomly with certain weights.

\begin{figure}
\hspace*{-0.5em}
\subfigure[]{
\includegraphics[width=0.51\linewidth]{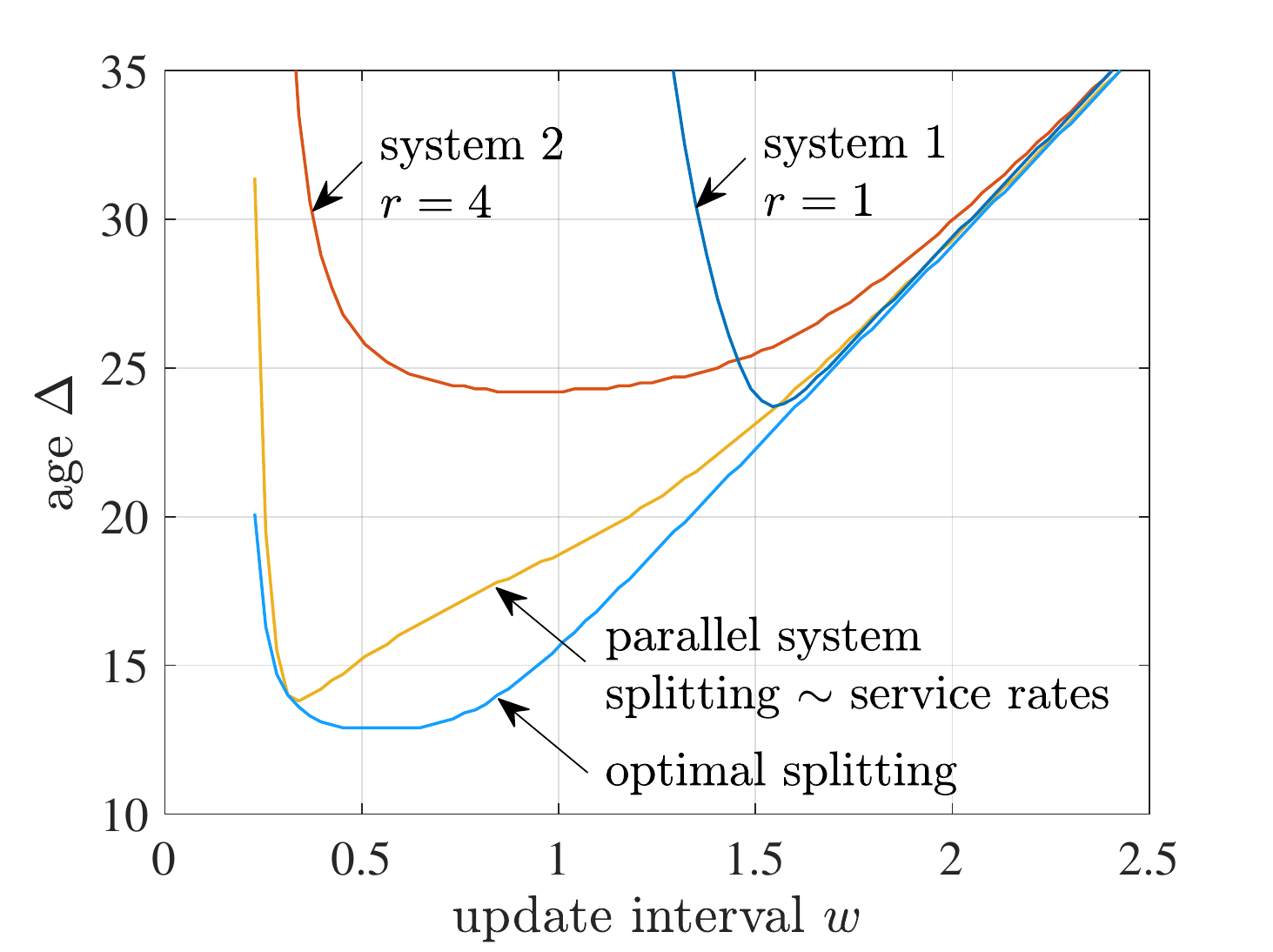}
\label{fig:onoffheterogeneous1}
}\hspace*{-1em}
\subfigure[]{
\includegraphics[width=0.51\linewidth]{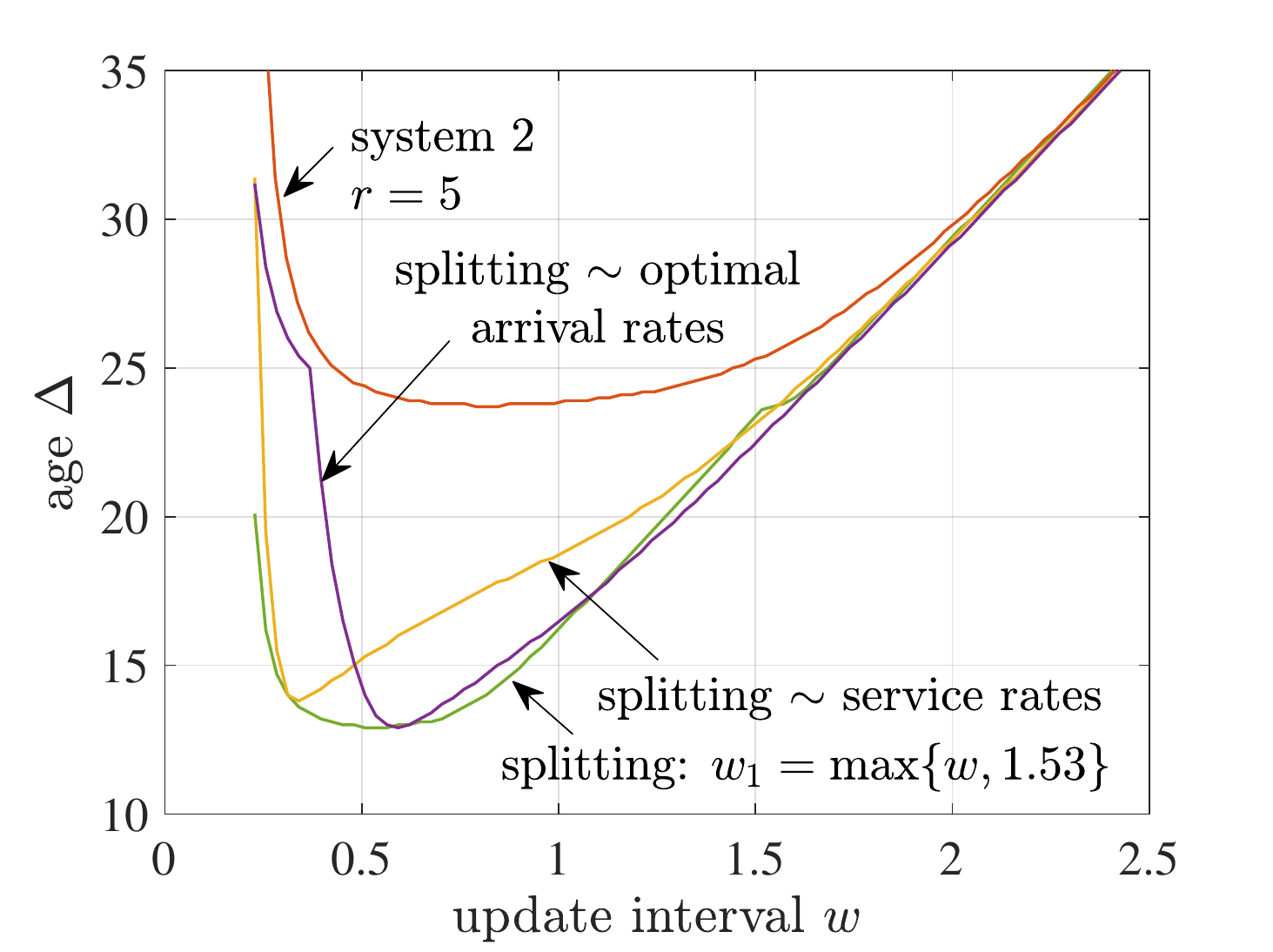}
\label{fig:onoffheterogeneous2}
}
\caption{Parallel system with two heterogeneous channels versus systems that use a single channel.}
\label{fig:onoffheterogeneous}
\end{figure}

Fig.~\ref{fig:onoffheterogeneous1} shows the age of the parallel system compared to the age of a system that uses either only channel 1 or channel 2, referred to as system 1 and system 2, respectively. Both, system 1 and system 2 have similar minimal age, despite the much lower mean service rate of system 1. This is due to the more frequent outages of system 2. In case of the parallel system, the arrivals are split randomly where the probabilities for splitting are proportional to the service rates, i.e., a ratio of 1 to 4. The parallel system can take advantage of the independence of the two channels, resulting in a significant reduction of the age. The minimal age of the parallel system is attained when the arrival rate $1/w$ is about $1/0.3$. For this operating point and the given splitting ratio of 1 to 4 we can notice that the arrival rate to channel 1 is about $1/1.53$, that is the arrival rate that minimizes the age of system 1. However, the arrival rate to channel 2 is about $1/0.38$ that is much larger than the update rate of $1/0.9$ that achieves system 2's minimal age. This motivates the investigation of other splitting weights. For reference, we include a curve of the age in Fig.~\ref{fig:onoffheterogeneous1} that is obtained as the optimal splitting by enumeration of all feasible ratios of random splitting.

Fig.~\ref{fig:onoffheterogeneous2} includes the age of two other random splitting policies: splitting proportionally to the optimal arrival rates of the single systems, i.e., a ratio of $1/1.53$ to $1/0.9$; and splitting so that channel 1 has a mean arrival rate of $1/1.53$, whenever possible, that is the arrival rate that minimizes the age of system 1. Both policies show an improvement of the minimal age. Further the minimal age is achieved with a lower update rate of about $1/0.6$, which saves resources.

In Fig.~\ref{fig:onoffheterogeneous2}, we also include the age of system 2 if we increase the mean service rate to 5, i.e., if system 2 has a mean service rate equal to the sum of the two channels used by the parallel system. However, due to the frequent outages of system 2, the age hardly benefits from the increased mean service rate. In contrast, we found that using heterogeneous channels in parallel is workable and can improve the age significantly. An effective splitting strategy will first consider the optimal update rate of the reliable, yet lower rate FR1 channel and assign remaining updates to the higher rate and less reliable FR2 channel.
%
%
\section{Related Works}
\label{sec:relatedworks}
Research works in age-of-information literature can be broadly divided into two categories: 1) analysis of age, and 2) minimization
of age. The latter category of works focused on designing optimal sampling and/or scheduling (for transmissions) for a range of status update systems for minimizing different statistics of age, e.g., see \cite{kaul:ageofinformationqueue,Yates2012,yates:lazyistimely,shroff:updateorwait,modiano:ageofinformationqueueing,Talak2019:ISIT1,Talak2019:ISIT2,Bedewy2019:TIT,Champati2021:JSAC}, and \cite{Yates2021:survey} for a comprehensive survey. Our work falls in the former category and we summarize the related works in the following.

Initial works on age analysis have focused on providing expressions for the average age and the peak age for different status-update systems, e.g., see \cite{kaul:ageofinformationqueue,Costa2016,Najm2016,Najm2017,Bedewy2019:TON,Soysal2021}. Later, the authors in \cite{inoue:aoisingleserverqueues,champati:ageofinformationgigiqueue,rizk:palmaoi} provided expressions for the distribution of age for single-source single-server systems. The authors in \cite{champati:ageofinformationmaxplus} used max-plus network calculus to derive CCDF bounds of the age in a tandem queueing network, while the authors in \cite{noroozi:minplusaoi} used min-plus network calculus to derive these bounds for different queueing settings.

In contrast to single-source single-server systems, works on the analysis of age in parallel server systems are relatively scarce. The authors in \cite{raiss:sensinghybridnetworks,altman:foreveryoung} studied the reception of messages (updates) by a mobile device user in a hybrid wireless network where the user can choose a cellular network or WiFi. Each received message is assigned a utility based on current age and a negative cost corresponding to the consumed energy and the monetary cost for the reception. They compute optimal scheduling policies that choose, in each slot, either to wait or use WiFi if available or use cellular data if WiFi is not available, in order to receive the messages. Motivated by the availability of millimeter-wave and sub-6 GHz channels in 5G, the authors in \cite{pan:hybridchannels} studied a single-source two parallel server system, where millimeter-wave was modelled by a Gilbert-Elliot channel and sub-6 GHz was modelled by a constant service time channel. Assuming that a packet can be served by only one channel at any time the authors computed optimal scheduling policies for minimizing the average age. In contrast to \cite{raiss:sensinghybridnetworks,altman:foreveryoung,pan:hybridchannels} we consider simultaneous transmissions on the parallel channels. In \cite{kam:agetransmissionpathdiversity}, the authors analyzed the average age and provided exact expressions for M$\mid$M$\mid$$\infty$ and an approximation for M$\mid$M$\mid$2 under FCFS discipline. In contrast to all the above works, we study the CCDF of the age and provide upper bounds for the random splitting policy under general service time distributions which include the Gilbert-Elliot channel.
%
%
\section{Conclusions}
\label{sec:conclusions}
Expressing the age-of-information of a parallel system as the minimum of the age of its subsystems raised the expectation that parallel systems are generally effective. An initial study of statistically independent parallel M$\mid$M$\mid$1 queues did not confirm this and it turned out that pooling the capacity into a single channel performs at least as good. To better understand the tradeoffs, we derived statistical age-of-information bounds for a broad class of parallel systems. A main finding of our work is that the effectiveness of parallel systems depends largely on the statistical properties of the subsystems. No advantage could be shown for parallel M$\mid$D$\mid$1, M$\mid$M$\mid$1, or D$\mid$M$\mid$1 queues. In contrast, parallel Markov channels have an edge over an equivalent single channel and the advantage becomes larger as the memory of the channels increases. This is because the age of a Markov channel with memory is dominated by long off periods of the channel that are effectively compensated for by the statistical independence of parallel channels. A significant benefit was also discovered for heterogeneous parallel channels as will occur for example in 5G multi-band communication. For scheduling policies, like join the shortest queue, we found that the additional effort may not be worthwhile, as they only have marginal impact on the age. Our results also confirm that periodic updates tend to be more practical as they reduce the age compared to Poisson updates in single as well as parallel systems.
%
%
\bibliographystyle{IEEEtran}
\bibliography{IEEEabrv,IEEEfidler}
%
%
\section{Appendix}
\label{sec:appendix}
\begin{proof}[{\bf Proof of Cor.~\ref{cor:packetizer}}]
From Def.~\ref{def:envelopes} we have for all $t \ge 0$ that
\begin{equation*}
\mathsf{P}[S(\tau,t) \ge [\rho_S (t-\tau) - b]_+ \; \forall \tau \in [0,t]] \ge 1-\varepsilon_S(b) .
\end{equation*}
For all $t \ge \tau \ge 0$ it can be concluded that if
\begin{align*}
& S(\tau,t) \ge \max \{0,\rho_S (t-\tau)-b\} \\
\Rightarrow & \max \{0,S(\tau,t)-l_{\max}\} \ge \max\{0,\rho_S (t-\tau)-l_{\max}-b\} .
\end{align*}
Hence, we have
\begin{equation*}
\mathsf{P}[S_{P_L}(\tau,t) \ge [\rho_S (t-\tau) - l_{\max} - b]_+ \; \forall \tau \in [0,t]] \ge 1-\varepsilon_S(b) .
\end{equation*}
where we inserted $S_{P_L}(\tau,t)$ from~\eqref{eq:servicepacketizer}. Taking the converse statement completes the proof.
\end{proof}
%
%
\begin{proof}[{\bf Proof of Th.~\ref{th:aoi}}]
Given the min-plus formulation of age~\eqref{eq:ageminplus} and with $A(t)$ non-decreasing and right-continuous we write
\begin{equation*}
\mathsf{P}[\Delta(t) > x] \le \mathsf{P}[\Delta(t) \ge x] = \mathsf{P}[A(t-x)-D(t) \ge 0] ,
\end{equation*}
for $t \ge x$. We substitute~\eqref{eq:serviceprocess} for $D(t)$ and use the service process with packetization~\eqref{eq:servicepacketizer} to obtain that
\begin{equation*}
A(t-x)-D(t) \le \sup_{\tau \in [0,t]} \{A(t-x) - A(\tau) - S_{P_L}(\tau,t)\} .
\end{equation*}
We divide $\sup_{\tau \in [0,t]} \{ . \} = \sup\{ \sup_{\tau \in [0,t-x]} \{ . \},\sup_{\tau \in [t-x,t]} \{ . \} \}$ and apply the union bound to estimate
\begin{align}
& \mathsf{P}[A(t-x)-D(t) \ge 0] \nonumber \\
& \le \mathsf{P} \biggl[ \sup_{\tau \in [0,t-x]}\{A(\tau,t-x) - S_{P_L}(\tau,t)\} \ge 0 \biggr] \label{eq:sup1}\\
& \, + \mathsf{P} \biggl[ \sup_{\tau \in [t-x,t],}\{-A(t-x,\tau)-S_{P_L}(\tau,t)\} \ge 0 \biggr] \label{eq:sup2} .
\end{align}
We use the approach~\cite[Chap. 6]{jiang:stochasticnetworkcalculus} and expand the first supremum in~\eqref{eq:sup1}:
\begin{align*}
& \sup_{\tau \in [0,t-x]}\{ A(\tau,t-x) - S_{P_L}(\tau,t)\} \\
= & \sup_{\tau \in [0,t-x]} \{ A(\tau,t-x) - \rho_A (t-x-\tau) + \rho_A (t-x-\tau) \\
& + \rho_S (t-\tau) - l_{\max} - S_{P_L}(\tau,t) - \rho_S (t-\tau) + l_{\max} \} \\
\le & Y_A + Y_S + \beta .
\end{align*}
Here, $Y_A = \sup_{\tau \in [0,t-x]} \{ A(\tau,t-x) - \rho_A(t-x-\tau) \}$ and with Def.~\ref{def:envelopes}
\begin{equation}
\mathsf{P} [Y_A > b] \le \overline{\varepsilon}_A(b) ,
\label{eq:za}
\end{equation}
$Y_S = \sup_{\tau \in [0,t-x]} \{ \rho_S (t-\tau) - l_{\max} - S_{P_L}(\tau,t) \}$ and with Cor.~\ref{cor:packetizer}
\begin{equation}
\mathsf{P} [Y_S > b] \le \varepsilon_S(b) ,
\label{eq:zs}
\end{equation}
and $\beta = \sup_{\tau \in [0,t-x]} \{ \rho_A (t-x-\tau) - \rho_S (t-\tau) + l_{\max} \}$ where $\rho_A \le \rho_S$ for stability and thus
\begin{equation*}
\beta = - \rho_S x + l_{\max} .
\end{equation*}
It follows for $\epsilon > 0$ that
\begin{align*}
& \mathsf{P} \biggl[ \sup_{\tau \in [0,t-x]}\{A(\tau,t-x) - S_{P_L}(\tau,t)\} \ge 0 \biggr] \\
\le & \mathsf{P} [Y_A + Y_S \ge -\beta ] \\
\le & \mathsf{P} [Y_A + Y_S > -\beta -\epsilon ] \\
\le & 1- [1-\overline{\varepsilon}_A]_+ \ast [1-\varepsilon_S]_+ (\rho_S x - l_{\max} - \epsilon) ,
\end{align*}
where we used that $Y_A$ and $Y_S$ are independent random variables and applied~\cite[Lem. 6.1]{jiang:stochasticnetworkcalculus}. Finally, we let $\epsilon \rightarrow 0$.

For the second sup~\eqref{eq:sup2} we use that $A(\tau,t)$ and $S_{P_L}(\tau,t)$ for $t \ge \tau \ge 0$ are non-negative and estimate
\begin{equation}
\mathsf{P} \biggl[ \inf_{\tau \in [t-x,t],}\{A(t-x,\tau)+S_{P_L}(\tau,t)\} \le 0 \biggr] \le \mathsf{P}[ Z_A + Z_S \ge x] .
\label{eq:sup2idletimes}
\end{equation}
Here, $Z_A = \sup \{ \tau \ge t-x: A(t-x,\tau) = 0 \} - (t-x)$ and with Def.~\ref{def:envelopes}
\begin{equation}
\mathsf{P}[Z_A > u] \le \underline{\varepsilon}_A(u) ,
\label{eq:ta}
\end{equation}
$Z_S = t - \inf \{\tau \in [0,t]: S_{P_L}(\tau,t) = 0\}$ and with Cor.~\ref{cor:packetizer} we have $\inf \{\tau \in [0,t]:  [\rho_S' (t-\tau) - l_{\max} - b]_+ = 0\} = t - (l_{\max} + b)/\rho_S'$ and
\begin{equation}
\mathsf{P}[Z_S > (l_{\max} + b)/\rho_S'] \le \varepsilon_S'(b) ,
\label{eq:ts}
\end{equation}
where $\rho_S'$ is a lower service envelope with underflow profile $\varepsilon_S'(b)$. With two variable substitutions $Z_S' = Z_S - l_{\max}/\rho_S'$ and $u = b/\rho_S'$
\begin{equation*}
\mathsf{P}[Z_S' > u] \le \varepsilon_S'(\rho_S' u) =: \varepsilon_T(u) .
\end{equation*}
By insertion into~\eqref{eq:sup2idletimes} it follows that
\begin{align*}
& \mathsf{P} \biggl[ \inf_{\tau \in [t-x,t],}\{A(t-x,\tau)+S_{P_L}(\tau,t)\} \le 0 \biggr] \\
\le & \mathsf{P}[ Z_A + Z_S' > x - l_{\max}/\rho_S - \epsilon ] \\
\le & 1- [1-\underline{\varepsilon}_A]_+ \ast [1-\varepsilon_T]_+ (x-l_{\max}/\rho_S'-\epsilon) ,
\end{align*}
where we used that $Z_A$ and $Z_S'$ are independent random variables and applied~\cite[Lem. 6.1]{jiang:stochasticnetworkcalculus}, as before. Finally, we let $\epsilon \rightarrow 0$.
\end{proof}
%
%
\begin{proof}[{\bf Proof of Lemma~\ref{lem:convolution}}]
Note that $\varepsilon_1(x) = 1$ at $x=\ln(\alpha)/\upsilon_1 > 0$ if $\alpha > 1$. For $x \ge [\ln(\alpha)]_+/\upsilon_1$ and $\upsilon_1 = \upsilon_2 =: \upsilon$ we have
\begin{align*}
& 1 - [1-\varepsilon_1]_+ \ast (1-\varepsilon_2) (x) \\
= & 1 - \int_0^{x}\bigl[1-\alpha e^{-\upsilon (x-y)}\bigr]_+ d\bigl(1- e^{-\upsilon y}\bigr) \\
= & 1 - \int_0^{x-\frac{[\ln \alpha]_+}{\upsilon}} \bigl(1-\alpha e^{-\upsilon (x-y)}\bigr) \bigl(\upsilon e^{-\upsilon y}\bigr) dy \\
= & 1 - \upsilon \int_0^{x-\frac{[\ln \alpha]_+}{\upsilon}} e^{-\upsilon y} dy + \alpha \upsilon e^{-\upsilon x} \int_0^{x-\frac{[\ln \alpha]_+}{\upsilon}} dy \\
= & e^{-\upsilon \bigl(x-\frac{[\ln \alpha ]_+}{\upsilon}\bigr)} + \alpha \upsilon e^{-\upsilon x} \biggl(x-\frac{[\ln \alpha]_+}{\upsilon}\biggr) \\
= & ([\alpha]_1 - \alpha [\ln \alpha]_+ + \alpha \upsilon x) e^{-\upsilon x} ,
\end{align*}
where $[\alpha]_1 = \max\{1,\alpha\}$. For $x \ge [\ln(\alpha)]_+/\upsilon_1$ and $\upsilon_1 \neq \upsilon_2$ we have
\begin{align*}
& 1 - [1-\varepsilon_1]_+ \ast (1-\varepsilon_2) (x) \\
= & 1 - \int_0^{x-\frac{[\ln \alpha]_+}{\upsilon_1}} \bigl(1-\alpha e^{-\upsilon_1 (x-y)}\bigr) \bigl(\upsilon_2 e^{-\upsilon_2 y}\bigr) dy \\
= & 1 - \upsilon_2 \int_0^{x-\frac{[\ln \alpha]_+}{\upsilon_1}} \!\!\!\! e^{-\upsilon_2 y} dy + \alpha \upsilon_2 e^{-\upsilon_1 x} \!\! \int_0^{x-\frac{[\ln \alpha]_+}{\upsilon_1}} \!\!\!\! e^{(\upsilon_1 - \upsilon_2) y} dy \\
= & e^{-\upsilon_2 \bigl(x-\frac{[\ln \alpha]_+}{\upsilon_1}\bigr)} + \frac{\alpha \upsilon_2 e^{-\upsilon_1 x}}{\upsilon_1-\upsilon_2} \biggl(e^{(\upsilon_1-\upsilon_2) \bigl(x-\frac{[\ln \alpha]_+}{\upsilon_1}\bigr)} - 1 \biggr) \\
= & [\alpha]_1^{\frac{\upsilon_2}{\upsilon_1}} e^{-\upsilon_2 x} + \frac{\alpha \upsilon_2}{\upsilon_1-\upsilon_2} \biggl([\alpha]_1^{\frac{\upsilon_2}{\upsilon_1}-1} e^{-\upsilon_2 x} - e^{-\upsilon_1 x} \biggr) .
\end{align*}
Some reordering completes the proof.
\end{proof}
%
%
\begin{proof}[{\bf Proof of Cor.~\ref{cor:constantrateservice}}]
The proof is a variation of the proof of Th~\ref{th:aoi}. For the first term~\eqref{eq:sup1} we find for $t \ge x$ and $x \ge l_{\max}/r$ that
\begin{align*}
& \mathsf{P} \biggl[ \sup_{\tau \in [0,t-x]}\{A(\tau,t-x) - S_{P_L}(\tau,t)\} \ge 0 \biggr] \\
\le & \mathsf{P}[Y_A \ge - \beta] \\
\le & \overline{\varepsilon}_A(r x - l_{\max}).
\end{align*}
Above $Y_A = \sup_{\tau \in [0,t-x]} \{ A(\tau,t-x) - \rho_A (t-x-\tau) \}$ satisfies \eqref{eq:za} and $\beta = \sup_{\tau \in [0,t-x]} \{ \rho_A (t-x-\tau) - [r (t-\tau)-l_{\max}]_+ \}$. With $\rho_A \le r$ we have $\beta = - r x + l_{\max}$ for $x \ge l_{\max}/r$.

For the second term~\eqref{eq:sup2} we use~\eqref{eq:sup2idletimes}. The random variable $Z_A$ satisfies~\eqref{eq:ta} as before and for constant rate service it follows that $Z_S = t - \inf \{\tau \in [0,t]:  [r (t-\tau) - l_{\max}]_+ = 0\} = l_{\max}/r$ for $t \ge l_{\max}/r$. By insertion into~\eqref{eq:sup2idletimes} we obtain that
\begin{align*}
& \mathsf{P} \biggl[ \sup_{\tau \in [t-x,t],}\{-A(t-x,\tau)-S_{P_L}(\tau,t)\} \ge 0 \biggr] \\
\le & \mathsf{P}[ Z_A \ge x - l_{\max}/r] \\
\le & \underline{\varepsilon}_A(x - l_{\max}/r) ,
\end{align*}
for $t \ge x$ and $x \ge l_{\max}/r$.
\end{proof}
%
%
\begin{proof}[{\bf Proof of Cor.~\ref{cor:periodicarrivals}}]
The proof is a variation of the proof of Th~\ref{th:aoi}. For the first term~\eqref{eq:sup1} we find for $t \ge x$ and $x \ge 2l/\rho_S$ that
\begin{align*}
& \mathsf{P} \biggl[ \sup_{\tau \in [0,t-x]}\{A(\tau,t-x) - S_{P_L}(\tau,t)\} \ge 0 \biggr] \\
\le & \mathsf{P}[Y_S \ge - \beta] \\
\le & \varepsilon_S(\rho_S (x - [l/\rho_S - (t-x+o) \bmod w]_+) - l).
\end{align*}
Above $Y_S = \sup_{\tau \in [0,t]} \{ \rho_S (t-\tau) - l_{\max} - S_{P_L}(\tau,t) \}$ satisfies \eqref{eq:zs} and $\beta = \sup_{\tau \in [0,t-x]} \{ A(\tau,t-x) - \rho_S (t-\tau) + l_{\max} \}$. With $A(t)$ given in Cor.~\ref{cor:periodicarrivals}, $\rho_S \ge l/w$ for stability, and $l_{\max} = l$ we derive $\beta = -\rho_S x + [l - \rho_S ((t-x+o) \bmod w)]_+ + l$.

For the second term~\eqref{eq:sup2} we use~\eqref{eq:sup2idletimes}. For periodic arrivals we have $Z_A = \sup \{\tau : A(t-x,\tau) = 0 \} - (t-x) = w - (t-x+o) \bmod w$ and the random variable $Z_S$ satisfies~\eqref{eq:ts} as before. By insertion into~\eqref{eq:sup2idletimes} we obtain
\begin{align*}
& \mathsf{P} \biggl[ \sup_{\tau \in [t-x,t],}\{-A(t-x,\tau)-S_{P_L}(\tau,t)\} \ge 0 \biggr] \\
\le & \mathsf{P}[ Z_S \ge x - w + (t-x+o) \bmod w] \\
\le & \varepsilon_S'(\rho_S' (x - (w - (t-x+o) \bmod w )) - l) ,
\end{align*}
for $t \ge x$ and $x \ge w+l/\rho_S'$.
\end{proof}
%
%
\begin{proof}[{\bf Derivation of the service envelope}]
We prove that~\eqref{eq:laplaceenvelope} and~\eqref{eq:serviceunderflowprofile} satisfy
\begin{equation}
\mathsf{P}[\exists \tau \in [0,t] : S(\tau,t) < \rho_S(\theta) (t-\tau) - b] \le e^{-\theta b} .
\label{eq:serviceenvelopeproof}
\end{equation}
Since $S(\tau,t)$ is non-negative, the service envelope in Def.~\ref{def:envelopes} follows from~\eqref{eq:serviceenvelopeproof}. We use the same steps as for the arrival envelope in~\cite{jiang:noteonsnetcalc}. For $\theta > 0$ we rewrite~\eqref{eq:serviceenvelopeproof} as
\begin{align*}
& \mathsf{P}\biggl[\sup_{\tau \in [0,t]} \{\rho_S(\theta) \tau - S(t-\tau,t) \} > b \biggr] \\
= & \mathsf{P}\Bigl[e^{\theta \sup_{\tau \in [0,t]} \{\rho_S(\theta) \tau - S(t-\tau,t) \}} > e^{\theta b} \Bigr] \\
= & \mathsf{P}\biggl[\sup_{\tau \in [0,t]} \Bigl\{ e^{\theta (\rho_S(\theta) \tau - S(t-\tau,t))} \Bigr\} > e^{\theta b} \biggr] .
\end{align*}
Now consider the process
\begin{equation*}
U(\tau)  = e^{\theta (\rho_S(\theta) \tau-S(t-\tau,t))} .
\end{equation*}
Given the service process has stationary and independent increments $S(\tau-1,\tau)$ where $S(0,t) = \sum_{\tau=1}^t S(\tau-1,\tau)$, it follows that
\begin{equation*}
U(\tau+1) = U(\tau) e^{\theta (\rho_S(\theta)-S(t-\tau-1,t-\tau))} ,
\end{equation*}
which has conditional expectation
\begin{align*}
& \mathsf{E}[U(\tau+1) | U(\tau), U(\tau-1),\dots, U(1)] \\
= & U(\tau) e^{\theta \rho_S(\theta)}\mathsf{E}\Bigl[e^{-\theta S(t-\tau-1,t-\tau)}\Bigr] \\
= & U(\tau),
\end{align*}
since $\rho_S(\theta) = -\frac{1}{\theta} \ln \mathsf{E}[e^{-\theta S(t-\tau-1,t-\tau)}]$. This means the process $U(\tau)$ is a martingale. With Doob's inequality and~\cite{jiang:noteonsnetcalc} we have for $U(\tau)$, $x$ non-negative, and all $t$
\begin{equation*}
\mathsf{P}\biggl[\sup_{\tau \in [1,t]} \{ U(\tau) \} > x \biggr] \le \frac{\mathsf{E}[U(1)]}{x} .
\end{equation*}
Finally, $\mathsf{E}[U(1)] = 1$ and $x = e^{\theta b}$ completes the proof.
\end{proof}
%
%
\noindent{\bf Numerical results for parallel M$\mid$E$_l$$\mid$1 queues}
\begin{figure}[h!]
\subfigure[M$\mid$E$_2$$\mid$1 queue]{
\includegraphics[width=0.46\linewidth]{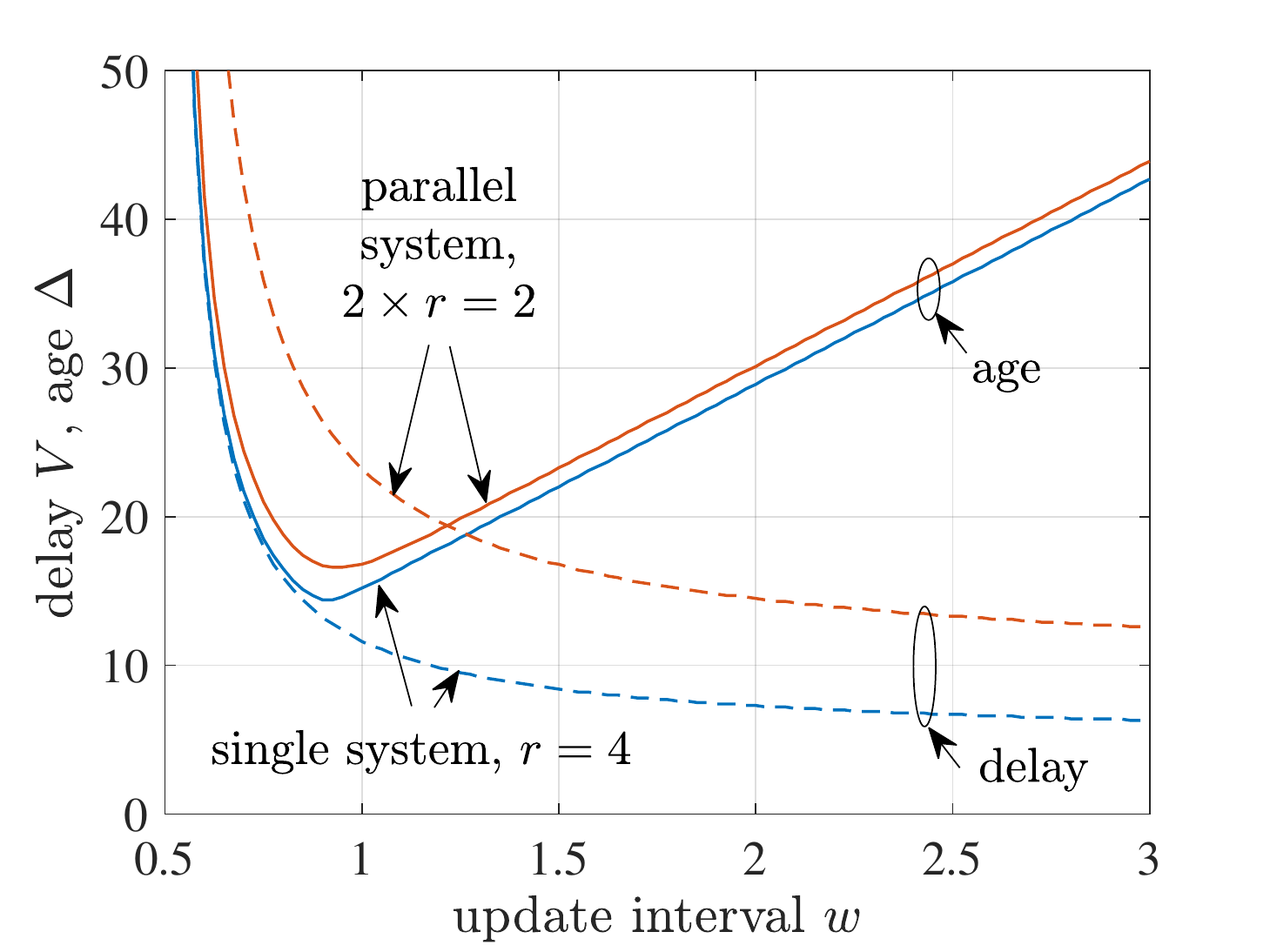}
\label{fig:me21_bound}
}
\subfigure[M$\mid$E$_2$$\mid$1 queue, simulation]{
\includegraphics[width=0.46\linewidth]{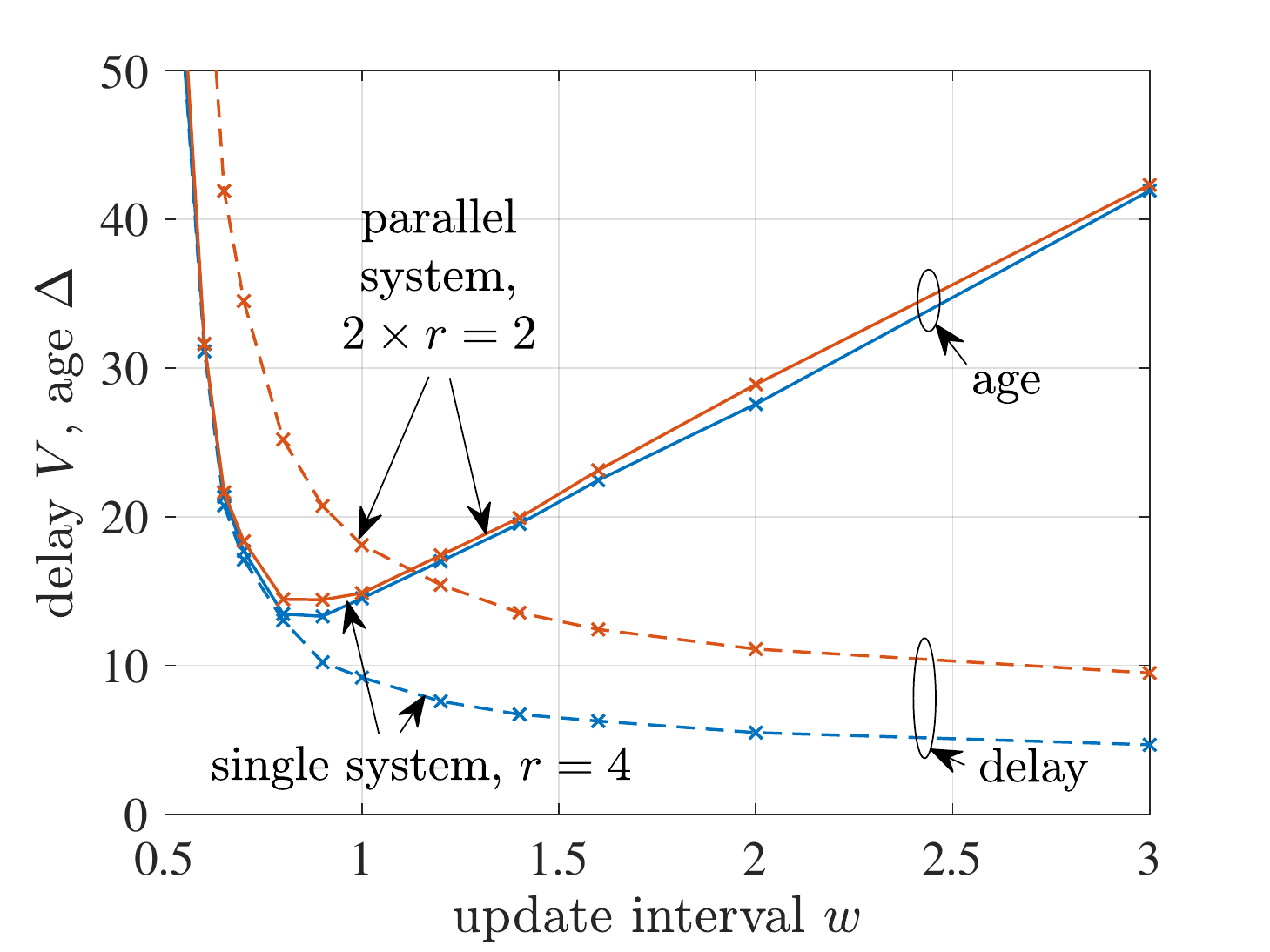}
\label{fig:me21_sim}
}
\caption{Single M$\mid$E$_l$$\mid$1 queue with service rate $r=4$ versus two parallel M$\mid$E$_l$$\mid$1 queues each with rate $r=2$, all for $l=2$. Compared to the M$\mid$M$\mid$1 queue, the minimal age and the update interval $w$ that minimizes the age are smaller. Due to the reduced variability of the service, M$\mid$E$_l$$\mid$1 queues tend to behave more like M$\mid$D$\mid$1 queues as $l$ and $r$ are increased proportionally.}
\label{fig:me21}
\end{figure}
\end{document}